# Rare-earth Bilayer Triangular Lattice as a Platform for Tunable Ground States and Field-induced Magnetic Phases

Jianqiao Wang[a,b,#], Fangli Li[c,#], Yangyang Yu[a,b], Quan Xiao[a,b], Zhibin Qiu[a,b], Liusuo Wu[c,*], Shu Guo[b,*]

[a]Southern University of Science and Technology, Shenzhen 518055, China.

[b]International Quantum Academy, Shenzhen 518048, China.

[c]Department of Physics, Southern University of Science and Technology, Shenzhen 518055, China.

[#]These authors contributed equally.
*Correspondence should be addressed to Liusuo Wu (wuls@sustech.edu.cn), and Shu Guo (shu-guo@outlook.com/shuguo@iqasz.cn).

**Abstract**

Rare-earth bilayer triangular-lattice (TL) antiferromagnets provide a versatile platform for realizing diverse magnetic states by combining geometric frustration, interlayer coupling, and strong single-ion anisotropy. Here, we report a family of $R_2O_2Se$ ($R$ = Sm, Eu, Tb–Lu; *R*OSe) single crystals featuring bilayer equilateral TLs. Magnetic susceptibility and specific-heat measurements reveal predominantly antiferromagnetic (AFM) interactions and diverse magnetic ground states across the series, including successive AFM transitions in TbOSe and single AFM transitions in SmOSe, DyOSe, HoOSe, and YbOSe. Notably, DyOSe and HoOSe exhibit pronounced 1/2 magnetization plateau-like features for fields along the *c*-axis, revealing field-induced magnetic states. In HoOSe, complementary thermodynamic and magnetization measurements resolve three critical fields and a rich field–temperature phase diagram containing five distinct magnetic phases. These results establish rare-earth bilayer TLs as a chemically tunable materials platform for accessing novel quantum spin states through the interplay of lattice geometry, single-ion anisotropy, and competing magnetic interactions.

## 1. Introduction

Geometric frustration arises primarily in lattices constructed from edge-, corner-, or face-sharing triangles or tetrahedra, where magnetic cations host the spins at the vertices of the quadrilaterals and triangles[1-3], with a 2D triangular lattice (TL) and a 3D pyrochlore as classic examples. [4-10] In geometric frustrated magnets (GFMs), magnetic ions are confined within a framework where antiferromagnetic (AFM) exchange interactions cannot be simultaneously satisfied[11,12], with the resulting competition and quantum fluctuations giving rise to complex and intriguing spin configurations[13,14]. Specifically, small spin quantum numbers lead to enhanced quantum fluctuations, which can promote the stabilization of exotic quantum phases such as quantum spin liquids (QSLs), typically observed in systems with $S = 1/2$ or $S = 1$ [6,15]. In QSLs, strong quantum fluctuations suppress conventional long-range magnetic order even at zero temperature, without breaking conventional symmetries[16,17]. Benefiting from these intrinsic properties, it might host topological order and fractionalized excitations, making it a promising platform for quantum computation and providing a key paradigm for unraveling the pairing mechanism of unconventional high-temperature superconductivity[18].

As one of the magnetic centers in the GFMs, rare earth ions ($R^{3+}$) can be generally classified into Kramers and non-Kramers ions based on the parity of their total angular momentum $J$ in general. Kramers $R^{3+}$ ions can usually be regarded as effective pseudospin-1/2 systems for the time-reversal symmetry-protected doublet ground state[7,19,20]. In contrast, in a non-Kramers $R^{3+}$ ion system, although a doublet ground state is not guaranteed, a pseudospin-1/2 could be realized for suitable crystalline electric fields (CEFs)[21]. Hence, $R$-based GFMs have been considered as a material platform for realizing exotic quantum spin states. For instance, the QSL candidates in pyrochlore oxides $Dy_2Ti_2O_7$ and $Tb_2Ti_2O_7$, as well as spin ice behavior in $Ho_2Ti_2O_7$, enrich the research landscape of GFMs[22-24]. $R$-based TL magnets, exemplified by $R$MgGaO$_4$[2,25] and $ARCh_2$ ($A$ = Na, K, Cs; $Ch$ = O, S, Se)[6,14,26], have been reported to be rich playgrounds for exploring novel quantum spin states, including QSLs and Berezinskii-Kosterlitz-Thouless phase[2,27]. Beyond unconventional quantum phases, *R*-based TL magnets also hold significant potential as magnetocaloric or regenerative materials, particularly those containing $Gd^{3+}$ or $Eu^{2+}$ [28-30]. Their vanishing orbital angular momentum, maximum spin ($S$ = 7/2), and correspondingly large magnetic entropy of R ln 8 per mole of magnetic ions [31-34]. Our previous work has demonstrated that the TL antiferromagnet $GdBO_3$, owing to its vanishing orbital angular momentum and large spin ($S$ = 7/2), exhibits excellent ultra-low-temperature magnetocaloric properties and has achieved 50 mK via a quasi-adiabatic demagnetization refrigeration device[35].

However, the above-mentioned materials are mainly single-layer TLs with relatively weak interlayer magnetic coupling. Incorporating double- or multi-layered TLs introduces additional degrees of freedom associated with interlayer exchange interactions and stacking configurations, enabling access to complex magnetic ground states and competing interactions in TL GFMs. For instance, the commercial regenerative materials $Gd_2O_2S$ (GOS) and its isostructural selenide analogue $Gd_2O_2Se$ (GOSe) were synthesized in polycrystalline or micrometer-sized single-crystal forms. Structurally, it exhibits equilateral bilayer TLs. Previous studies have demonstrated that these materials possess excellent luminescence properties[36,37]. In terms of the magnetic properties, fundamental properties of polycrystalline *R*OSe (*R* = Gd, Tb, Dy, Ho, Er, Tm) were first reported by G. Quezel[38-40]. Although the elementary magnetism of *R*OSe (*R* = Gd-Tm) has been reported previously.[38] However, owing to the difficulty in obtaining *R*OSe single crystals, its anisotropic magnetic properties as well as the non-negligible interlayer coupling within the bilayer TLs remain unexplored[38].

In this work, millimeter-scale *R*OSe (*R* = Sm, Eu, and Tb-Lu) single crystals were successfully grown for the first time via the chemical vapor transport (CVT) method. On this basis, their structure, magnetism, and specific heat were systematically investigated using single crystals. The results show that the two neighboring TLs of $R^{3+}$ ions forming bilayer TLs are arranged in an AB stacking configuration, with adjacent bilayers separated by intervening Se layers. And the variation in the number of 4*f* electrons induced by $R^{3+}$ ions, together with the unique bilayer TLs stacking structure, endows *R*OSe with abundant dynamic magnetism. Among the *R*OSe series, HoOSe provides a model platform for exploring the interplay between strong Ising-type single-ion anisotropy of the non-Kramers $Ho^{3+}$ and geometric frustration in a bilayer TL, which gives rise to field-induced metamagnetic transitions and intermediate plateau-like magnetic states.

## 2. Results and discussion

### 2.1 Crystal structure

Grown via the CVT method, *R*OSe (*R* = Sm, Eu, and Tb-Lu) single crystals are shown in **Figure 1**. The series of *R*OSe materials is characterized by single-crystal X-ray diffraction (SC-XRD) with the *P*-3*m*1 space group at 100 K (**Figures 1A-1D**). Each $R^{3+}$ ion is coordinated by four O and three Se atoms to form a seven-coordinate polyhedron described as a one-capped distorted trigonal antiprism, which is edge-sharing within the *ab*-plane and stacked along the *c*-axis[36]. This edge-sharing polyhedral connectivity is separated by $Se^{2-}$ ion layers, forming a spaced bilayer TL of $R^{3+}$ ions in the *ab*-plane. Considering the isostructural nature of these materials, the Dy version was used as a representative in the discussion. In the *ab*-plane, $Dy^{3+}$ ions form a regular TL with an in-plane nearest-neighbor (NN) distance of 3.83 Å. The second layer of the $Dy^{3+}$ TL is stacked in an alternating sequence, with a relatively larger interlayer distance of 3.61 Å within the bilayer (**Figure 1D**), and the distance between adjacent bilayers is 4.52 Å (**Figure 1A**). Intriguingly, within each bilayer structure, the TLs of the two layers are mutually displaced with respect to one another, as illustrated in **Figure 1B**. This displacement causes the *R* ions in one layer to project onto the triangular centers formed by *R* ions of the adjacent layer along the *c*-axis. Such structural features are analogous to those previously reported for $R_2O_2CO_3$[41]. Owing to the relatively short spacing within each bilayer, the *R*OSe structure can also be regarded as a TL system composed of alternating layers with inequivalent interlayer separations. **Figure 1F** exhibits the crystal morphology of *R*OSe, which matched well with their crystal symmetry. The crystallographic parameters of the materials are presented in **Tables S1** and **S2**. Lanthanide contraction is reflected in the systematic decrease in lattice parameters with increasing atomic number (**Figure S1**), and the energy dispersive spectroscopy distributions of *R* and Se elements further corroborate the structural integrity (**Figure S2**).

### 2.2 Heat capacity

The $R^{3+}$ ions with distinct magnetic moments, stemming from their diverse electronic configurations, exhibit different specific heat characteristics, as depicted in **Figures 2A-2I**. $\lambda$-shaped specific peaks, which are associated with long-range magnetic orderings, are observed in the Sm-, Tb-, Dy-, Ho-, Er-, and Yb-based compounds. Notably, TbOSe exhibits two distinct specific-heat peaks at 4.95 and 2.39 K (**Figure 2C**), respectively, indicative of complex magnetic phase transitions. In the case of HoOSe, a $\lambda$-transition peak appears at a $T_{N(Ho)}$ = 3.62 K, and broad humps emerge around 0.6 K and 20 K (**Figure 2E**). The broad hump around 0.6 K likely reflects additional low-temperature magnetic correlations, whereas the broad feature around 20 K may be attributed to the thermal population of CEF levels of $Ho^{3+}$ ions. With a further increase in the quantity of 4*f* electrons, the $\lambda$-transition peak is no longer observed for ErOSe and TmOSe above 2 K, though a corresponding phase transition remains present at 0.89 K for ErOSe. Broad humps were observed around 12 and 11 K for ErOSe and TmOSe, respectively (**Figures 2F** and **2G**). For $4f^{13}$, as shown in **Figure 2H**, YbOSe exhibits AFM transition around 1.76 K, which is captured in the zero-field specific heat. EuOSe has a nonmagnetic singlet ground state and exhibits no magnetic transition, consistent with the Van Vleck paramagnetism expected for $Eu^{3+}$ (**Figure 2B**). The specific heat of LuOSe can serve as a reliable phonon reference for other compounds due to the absence of magnetic contribution overlap (**Figure 2I**). After subtracting the lattice contributions, the magnetic specific heat $C_M/T$ of *R*OSe is shown in **Figure S3**, and the corresponding magnetic entropies are shown in **Figure S4**. For most magnetic members of *R*OSe, the magnetic entropy approaches approach (R ln 2) at low temperatures, indicating an effective two-state low-energy manifold. For the Kramers ions, the doublet degeneracy is protected by Kramers theorem, consistent with the Ising-type doublet ground state previously established for DyOSe by neutron diffraction and crystal field analysis[42]. For non-Kramers ions, however, the degeneracy is not symmetry-protected and is therefore more sensitive to local CEFs. For the non-Kramers ions, previously reported crystal field calculations on $Tb_2O_2S$ show that multiple low-lying levels, highlighting the important role of CEF effects in determining the low-energy magnetic degrees of freedom of $Tb^{3+}$[39].

### 2.3 Magnetic Susceptibility and Field-Dependent Magnetization.

To clarify the anisotropic magnetic behavior, direct-current (DC) magnetic susceptibility measurements were performed on as-grown *R*OSe crystals with different applied magnetic field directions. Based on magnetic anisotropy, the *R*OSe family can be classified into three categories: (i) weak anisotropy in TbOSe (**Figure 3A**), as noted in previous work. Under CEF, the magnetic moments of ground state Tb ions adopt a tilted configuration relative to the *c*-axis[43]; (ii) easy-axis anisotropy along the crystallographic *c*-axis is found in DyOSe (**Figure 3B**) and HoOSe (**Figure 3C**); (iii) easy-plane magnetic anisotropy for Er-, Tm-, and Yb-based compounds (**Figures 3D-F**). The magnetic transitions identified from the susceptibility data are consistent with the specific heat peaks and are reproduced under applied magnetic fields along different directions (**Figures S5-S6**).

As shown in **Figure 3A**, TbOSe exhibits two consecutive transitions under an applied magnetic field of 0.05 T, consistent with the two distinct peaks observed in specific heat, in which the lower AFM transition has remained unaddressed in earlier polycrystalline-based investigations[39]. In contrast, AFM orderings were observed in DyOSe and HoOSe. Notably, HoOSe features a broad hump around 20 K, which matches the corresponding feature in the specific heat. The long-range magnetic ordering is absent in either TmOSe or ErOSe for both orientations above 1.8 K, consistent with the specific heat measurements. Zero-field-cooling (ZFC) and field-cooling (FC) comparison curves for *R*OSe materials show good overlap, except for Sm (**Figures S5 and S6**). The weak bifurcation between the ZFC and FC susceptibilities below the ordering temperature suggests potential magnetic irreversibility, possibly associated with domain effects or weak spin canting (**Figure S5A**). Supplementary field-dependent specific-heat measurements were therefore performed on the SmOSe compound (**Figure S7**). Under applied magnetic fields ranging from 0 to 9 T, neither suppression nor a shift of the specific-heat peak was observed, which is uncommon for a conventional AFM transition and is consistent with the magnetization behavior (**Figure S8**). The well-defined Curie-Weiss (C-W) behaviors could be inferred from the $1/\chi$ curves for most members in the *R*OSe family. As shown in **Figures S9**, **S10**, and **Table S3**, linear C-W fits yield negative values for most compounds in *R*OSe, indicating dominant AFM interactions[26,41]. As shown in **Figures 4A-4C**, field-dependent magnetization measurements reveal transitions resembling spin-flop behavior in TbOSe, DyOSe, and HoOSe. In the TbOSe (**Figure 4A**), two metamagnetic transitions were observed at approximately 1.0 and 2.2 T, respectively, independent of crystal orientation. Interestingly, a pronounced 1/2 magnetization plateau-like feature was found in the DyOSe and HoOSe when the magnetic field was applied along the *c*-axis. The maximum magnetization at the highest magnetic fields is summarized in **Table S3**. Unlike the 1/3 magnetization plateau induced by the Up-Up-Down phase commonly found in TL antiferromagnets, the AB-stacked bilayer TLs introduce multiple competing energy scales[44,45]. $R_2O_2CO_3$ adopts analogous bilayer TL crystal structures; however, no clear features of magnetization plateaus have been observed[40,41,43,46]. Structurally, the large interbilayer separation (> 4.5 Å) of *R*OSe, compared with the short intrabilayer *R*–*R* distances (~3 Å), is expected to favor predominantly intrabilayer magnetic correlations, thereby providing a structural basis for the complex magnetic behavior observed in *R*OSe. This suggests that the 1/2 plateau-like feature may arise from competing magnetic interactions involving interlayer coupling within the bilayer TLs. Together with the strong Ising-type anisotropy along the *c*-axis, these competing interactions may stabilize an intermediate field-induced magnetic state with approximately half of the saturation magnetization[44]. The microscopic spin configuration underlying this state, however, remains to be established.

### 2.4. Magnetic Phase Diagram of HoOSe

In this context, HoOSe is particularly useful for disentangling the field evolution of the magnetic states. While the 1/2 plateau-like feature is observed in both DyOSe and HoOSe along the *c*-axis, the successive field-induced phases in HoOSe can be further mapped using multiple thermodynamic and magnetic signatures. Accordingly, a specific-heat magnetic phase diagram is presented in **Figure 5**, constructed from temperature-dependent $C_p(T)$, field-dependent $C_p(B)$, magnetocaloric effect (MCE), and differential magnetization $dM/dB$ measurements, which reveal a rich sequence of successive magnetic phase transitions. At zero field, the system enters a long-range AFM phase (**I**) upon cooling below 3.62 K, while an anomaly is observed at 0.52 K. This long-range AFM phase is suppressed under an applied magnetic field, where an AFM phase (**I**) to disordered phase (**IV**) occurs at $B$ = 1.25 T. The phase boundaries are determined from the anomalies and extrema of $C_p(T)$, $C_p(B)$, and MCE measurements, enabling the identification of two additional phases emerging upon cooling within this field window: phase (**II**) ($B$ = 1.25 T, $T$ = 1.75 K), phase (**III**) ($B$ = 1.85 T, $T$ = 1.44 K). These phases are separated by critical fields $B_{c1}$ = 0.82 T, $B_{c2}$ = 1.88 T, and $B_{c3}$ = 2.97 T. Beyond $B_{c3}$, the fully polarized phase (**V**) is stabilized, with a crossover forming the boundary between phase (**IV**) and phase (**V**). **Figure S11** illustrates the magnetization behavior in the low temperature regime below 1.8 K. Broad high intensity regions are observed near 1 T and 3 T, which correspond to the rapidly rising regimes of magnetization. Upon varying the temperature, a new transition region emerges near 1.8 T. These three regions are associated with the phase transitions in the specific heat phase diagram, marking the magnetic field range where spin correlations develop and coincide with the critical fields. As shown in **Figure S12**, two distinct features are observed at 1/3 and 1/2 of the saturation magnetization at 0.6 K, which confirms the existence of thermally stabilized intermediate-field phases (**II**) and (**III**). As shown in **Figure S13A**, the low-field susceptibility curve measured at 0.05 T exhibits a weak broad feature around 0.5 K, suggesting the development of additional low-temperature magnetic correlations. In-plane susceptibility measurements along, for instance, the *a* and *a** directions reveal no pronounced anisotropic evolution in the HoOSe compound, as depicted in **Figure S13B-13C**. Further studies combining single-crystal neutron experiments, nuclear magnetic resonance, and muon spin rotation/relaxation measurements are needed to determine the magnetic structures and elucidate the low-energy spin dynamics of HoOSe and other members of the *R*OSe family.

## Conclusion

In summary, we have successfully grown millimeter-sized single crystals of *R*OSe (*R* = Sm-Eu and Tb-Lu) via chemical vapor transport. The unique AB-stacked bilayer TLs structure introduces new degrees of freedom and complex competing interlayer and intralayer interactions. AFM orderings were observed in the Sm-, Tb-, Dy-, Ho-, and Yb-based compounds based on specific heat and magnetic susceptibility analyses. Furthermore, two successive metamagnetic transitions are observed in the Tb-based compound, showing no discernible magnetic-field-direction dependence. Notably, a fractional 1/2 plateau-like feature emerges in the Dy- and Ho-based compounds under magnetic fields applied along the crystallographic *c*-axis. Furthermore, the magnetic phase diagram of HoOSe reveals a rich landscape of field-induced intermediate magnetic states and their evolution with temperature and magnetic field. Collectively, our results demonstrate how rare-earth single-ion characteristics, magnetic interactions, and bilayer geometry cooperate to generate diverse magnetic ground states and field-induced phases in the *R*OSe family.

**Methods**

**Single crystal growth**: Referring to past literature[34,36,47], the single crystals of *R*OSe were prepared via the reaction:

$$R_2O_3/0.5Tb_4O_7 + Se + 0.5C/0.75C \xrightarrow{I_2,CsCl} R_2O_2Se + 0.5CO_2/0.75CO_2$$

High-purity $R_2O_3$, $Tb_4O_7$, Se, and C were weighed and uniformly ground in an agate mortar. $R_2O_3$, Se, and C were weighed in a molar ratio of 2:4:1, whereas $Tb_4O_7$, Se, and C were mixed in a molar ratio of 1:4:1 for the Tb-based sample. Trace amounts of $I_2$ and CsCl (around three times the molar amount of $R_2O_3$) were added to a quartz ampoule, which was then evacuated and sealed. The ampoule was heated to 1000 ℃ in a muffle furnace and held for 48 hours, followed by slow cooling to 700 ℃ inside the furnace and further gradual cooling to room temperature. The product was washed with distilled water to remove cesium chloride and dried in air, yielding transparent hexagonal single crystals.

**SC-XRD:** SC-XRD experiments were conducted on a Bruker D8 VENTURE diffractometer equipped with a PHOTON III CPAD detector. The experiments were performed at 100 K in a temperature-controlled system utilizing nitrogen. X-rays ($\lambda$ = 0.71073 Å) were generated using a graphite monochromated Mo target. Background, polarization, and Lorentz factor corrections were subsequently applied using the APEX4 software, and multi-scan absorption correction was carried out with the SADABS package. The direct method of the ShelXT program was used for the structural solution, and the ShelXL least squares refinement package of Olex2 software was used for structural refinement[48,49]. The crystal structure was subjected to X-ray simulation using the Single Crystal 5 software package, yielding weighted diffraction points with structure factors[50].

**Elemental analysis:** Elemental composition analysis was carried out on a ZEISS GeminiSEM 300 scanning electron microscope (SEM) equipped with an Oxford energy dispersive spectroscopy using AZtecOne software.

**Magnetic measurements:** Direct-current (DC) magnetic susceptibility was measured using Physical Property Measurement System (PPMS) and the magnetic property measurement system (MPMS3) from Quantum Design, with the Vibrating Sample

Magnetometer (VSM) option. Based on a well-shaped single crystal piece, the magnetic field direction was adjusted to be parallel to the TL layer (in-plane magnetic field or $B \perp c$) and perpendicular to the TL layer (out-of-plane magnetic field or $B//c$) for testing. The low-temperature magnetic data were measured using a Quantum Design MPMS3, with data below 1.8 K obtained with the He-3 option; the testing mode was DC. Small single-crystal samples weighing 0.68 mg were measured, with the magnetic field oriented perpendicular to the TL layer.

**Specific heat measurements:** The specific heat was measured on a PPMS using the adiabatic heat-pulse method, with the applied magnetic field aligned along the *c*-axis ($B//c$) for all samples. Owing to potential mass uncertainties associated with tiny single crystals used in low-temperature specific heat experiments, data acquired from 0.4-1.8 K were corrected using specific heat values measured over the 1.8-150 K range. A dilution refrigerator (DR) module integrated into the PPMS platform enabled low-temperature characterization of the Ho-based sample over the 0.1-4 K range. Two single crystals of different masses were employed for the measurements: a 0.1 mg sample was used for the 0.4-1.1 K interval, whereas a 0.3 mg sample was measured from 1.1-4 K. Magnetocaloric effect (MCE) measurements were carried out by sweeping the magnetic field while tracking the sample temperature via a calibrated Cernox sensor (Lake Shore). Measurements were performed on a small single-crystal sample with the field applied along the *c* axis, using a homemade sample puck in a PPMS. The sample was mounted on a sapphire plate supported by a fiberglass frame, and the puck was installed in the high-vacuum chamber of a DR. To ensure quasi-adiabatic conditions, the sample platform was thermally isolated from the cold bath. At a constant bath temperature, the magnetic field was swept at a rate of 10-30 Oe/s, and the resulting change in sample temperature during the sweep was recorded as the MCE signal.

## Supporting Information

Extended data of lanthanide contraction, crystallographic information, EDS analysis, anisotropic magnetic susceptibility, Curie–Weiss fitting and associated magnetic parameters, and heat-capacity data.

[CCDC numbers 2588400(Dy), 2588401(Er), 2588402(Eu), 2588403(Ho), 2588404(Lu), 2588405(Sm), 2588406(Tb), 2588407(Tm), and 2588408(Yb) contain the supplementary crystallographic data for *R*OSe under 100 K, respectively. These data can be obtained free of charge from The Cambridge Crystallographic Data Centre via www.ccdc.cam.ac.uk/data_request/cif.]

## Acknowledgments

The authors acknowledge the financial support from the National Natural Science Foundation of China (22205091, 12374146) and the Guangdong Pearl River Talent Plan (2023QN10C793), the National Key Research and Development Program of China (Grant No. 2021YFA1400400, 2025YFA1411503), the Guangdong Basic and Applied Basic Research Foundation (Grant No. 2024B1515120045), the Guangdong Provincial Quantum Science Strategic Initiative (Grant No. GDZX2401006).

J. Q. Wang and F. L. Li contributed equally to this work.

## Declaration of interests

The authors declare no conflict of interest.

## Data Availability Statement

The data that support the findings of this study are available from the corresponding author upon reasonable request (shu-guo@outlook.com/shuguo@iqasz.cn).

## Reference

1. Lei, Z., Sathish, C.I., Geng, X., Guan, X., Liu, Y., Wang, L., Qiao, L., Vinu, A., and Yi, J. (2022). Manipulation of ferromagnetism in intrinsic two-dimensional magnetic and nonmagnetic materials. *Matter* 5, 4212–4273. 10.1016/j.matt.2022.11.017.
2. Liu, C., Huang, C.-J., and Chen, G. (2020). Intrinsic quantum Ising model on a triangular lattice magnet $TmMgGaO_4$. *Phys. Rev. Research* 2, 043013.. 10.1103/PhysRevResearch.2.043013.
3. Chen, J., Calder, S., Paddison, J.A., Angelo, G., Klivansky, L., Zhang, J., Cao, H., and Gui, X. (2025). $ASb_3Mn_9O_{19}$ (A= K or Rb): New Mn-Based 2D Magnetoplumbites with Geometric and Magnetic Frustration. *Adv. Mater.* 37, 2417906. 10.1002/adma.202417906.
4. Ramirez, A.P. (1994). Strongly geometrically frustrated magnets. *Annu. Rev. Mater. Sci.* 24, 453–480. 10.1146/annurev.ms.24.080194.002321.
5. Guo, R., Li, F., Zhao, N., Li, B., Xiao, Q., Zhang, C., Qiu, Z., Wang, J., Li, Z., and Wang, X. (2026). A Spin-5/2 Triangular-Lattice Antiferromagnet Exhibiting Field-Driven Competing Magnetic Phases. *J. Am. Chem. Soc.* 148, 20497–20508. 10.1021/jacs.6c00472.
6. Xing, J., Sanjeewa, L.D., Kim, J., Stewart, G.R., Du, M.-H., Reboredo, F.A., Custelcean, R., and Sefat, A.S. (2019). Crystal Synthesis and Frustrated Magnetism in Triangular Lattice $CsRESe_2$ (RE = La–Lu): Quantum Spin Liquid Candidates $CsCeSe_2$ and $CsYbSe_2$. *ACS Mater. Lett.* 2, 71–75. 10.1021/acsmaterialslett.9b00464.
7. Gao, Y., Xu, L., Tian, Z., and Yuan, S. (2018). Synthesis and magnetism of $RE(BaBO_3)_3$ (*RE* = Dy,Ho,Er,Tm,Yb) series with rare earth ions on a two dimensional triangle-lattice. *J. Alloys Compd.* 745, 396–400. 10.1016/j.jallcom.2018.02.110.
8. Lin, W., Li, Z., Zhang, C., Guo, R., Xiao, Q., An, W., Wen, B., Wang, Z., Sheng, J., and Wu, L. (2026). A $Mn^{2+}$-Based Pyrochlore Magnet With Fragile Magnetic Order and Strong Spin Fluctuations. *Adv. Mater. 38*, e21218. 10.1002/adma.202521218.
9. Guo, R., Sheng, J., Li, B., Guo, K., Yao, W., Zhao, N., Wang, J., Qiu, Z., Wen, B., and Ji a, S. (2026). Towards equilateral triangular lattice frustrated quantum magnets through crystal symmetry–protected molecular-brick chemical strategy. *Chin. Chem. Lett.* 37, 112623. 10.10 16/j.cclet.2026.112623.
10. Liu, J., Ding, Y., Zeng, M., and Fu, L. (2022). Chemical insights into two-dimensional qua ntum materials. Matter 5, 2168–2189. 10.1016/j.matt.2022.05.034.
11. Guo, S., Krug, D.A., Billingsley, B.R., Wang, J., Qiu, Z., and Kong, T. (2025). Magnetic compounds with exotic Archimedean lattices. *Innovation*. 5, 100981. 10.1016/j.xinn.2025.100981.
12. Fan, C., Chang, T., Fan, L., Teat, S.J., Li, F., Feng, X., Liu, C., Wang, S., Ren, H., Hao, J., et al. (2025). Pyrochlore $NaYbO_2$: A Potential Quantum Spin Liquid Candidate. *J. Am. Chem. Soc.* 147, 5693–5702. 10.1021/jacs.4c13166.
13. Damle, K. (2015). Melting of Three-Sublattice Order in Easy-Axis Antiferromagnets on Triangular and Kagome Lattices. *Phys. Rev. Lett.* 115, 127204. 10.1103/PhysRevLett.115.127204.
14. Xing, J., Taddei, K.M., Sanjeewa, L.D., Fishman, R.S., Daum, M., Mourigal, M., dela Cruz, C., and Sefat, A.S. (2021). Stripe antiferromagnetic ground state of the ideal triangular lattice compound $KErSe_2$. *Phys. Rev. B.* 103, 144413. 10.1103/PhysRevB.103.144413.
15. Li, Y., Chen, G., Tong, W., Pi, L., Liu, J., Yang, Z., Wang, X., and Zhang, Q. (2015). Rare-Earth Triangular Lattice Spin Liquid: A Single-Crystal Study of $YbMgGaO_4$. *Phys. Rev. Lett.* 115, 167203. 10.1103/PhysRevLett.115.167203.
16. Dun, Z.L., Trinh, J., Li, K., Lee, M., Chen, K.W., Baumbach, R., Hu, Y.F., Wang, Y.X., Choi, E.S.,

Shastry, B.S., et al. (2016). Magnetic Ground States of the Rare-Earth Tripod Kagome Lattice $Mg_2RE_3Sb_3O_{14}$(RE=Gd, Dy, Er). Phys. Rev. Lett. *116*, 157201. 10.1103/PhysRevLett.116.157201.

17. Anderson, P.W. (1973). Resonating valence bonds: A new kind of insulator? *Mater. Res. Bull*. 8, 153–160. 10.1016/0025-5408(73)90167-0.
18. Wen, J., Yu, S.-L., Li, S., Yu, W., and Li, J.-X. (2019). Experimental identification of quantum spin liquids. *npj Quantum Mater.* 4, 12. 10.1038/s41535-019-0151-6.
19. Ashtar, M., Gao, Y.X., Wang, C.L., Qiu, Y., Tong, W., Zou, Y.M., Zhang, X.W., Marwat, M.A., Yuan, S.L., and Tian, Z.M. (2019). Synthesis, structure and magnetic properties of rare-earth $REMgAl_{11}O_{19}$ (RE = Pr, Nd) compounds with two-dimensional triangular lattice. *J. Alloys Compd.* 802, 146–151. 10.1016/j.jallcom.2019.06.177.
20. Sanders, M.B., Krizan, J.W., and Cava, R.J. (2016). $RE_3Sb_3Zn_2O_{14}$ (RE = La, Pr, Nd, Sm, Eu, Gd): a new family of pyrochlore derivatives with rare earth ions on a 2D Kagome lattice. J. Mater. Chem. C. 4, 541–550. 10.1039/c5tc03798k.
21. Iwasa, K., Kobayashi, H., Onimaru, T., T. Matsumoto, K., Nagasawa, N., Takabatake, T., Ohira-Kawamura, S., Kikuchi, T., Inamura, Y., and Nakajima, K. (2013). Well-Defined Crystal Field Splitting Schemes and Non-Kramers Doublet Ground States of f Electrons in $PrT_2Zn_{20}$ (T= Ir, Rh, and Ru). *J. Phys. Soc. Jpn.* 82, 043707. 10.7566/JPSJ.82.043707.
22. Fennell, T., Deen, P.P., Wildes, A.R., Schmalzl, K., Prabhakaran, D., Boothroyd, A.T., Aldus, R.J., McMorrow, D.F., and Bramwell, S.T. (2009). Magnetic Coulomb Phase in the Spin Ice $Ho_2Ti_2O_7$. *Science* 326, 415–417. 10.1126/science.1177582.
23. Morris, D.J.P., Tennant, D.A., Grigera, S.A., Klemke, B., Castelnovo, C., Moessner, R., Czternasty, C., Meissner, M., Rule, K.C., Hoffmann, J.-U., et al. (2009). Dirac Strings and Magnetic Monopoles in the Spin Ice $Dy_2Ti_2O_7$. *Science* 326, 411–414. 10.1126/science.1177868.
24. Yaouanc, A., Dalmas de Réotier, P., Chapuis, Y., Marin, C., Vanishri, S., Aoki, D., Fåk, B., Regnault, L.P., Buisson, C., Amato, A., et al. (2011). Exotic transition in the three-dimensional spin-liquid candidate $Tb_2Ti_2O_7$. *Phys. Rev. B.* 84, 184403. 10.1103/PhysRevB.84.184403.
25. Shen, Y., Liu, C., Qin, Y., Shen, S., Li, Y.-D., Bewley, R., Schneidewind, A., Chen, G., and Zhao, J. (2019). Intertwined dipolar and multipolar order in the triangular-lattice magnet $TmMgGaO_4$. *Nat. Commun.* 10, 4530. 10.1038/s41467-019-12410-3.
26. Dissanayaka Mudiyanselage, R.S., Wang, H., Vilella, O., Mourigal, M., Kotliar, G., and Xie, W. (2022). $LiYbSe_2$: Frustrated Magnetism in the Pyrochlore Lattice. *J. Am. Chem. Soc.* 144, 11933–11937. 10.1021/jacs.2c02839.
27. Shen, Y., Li, Y.D., Wo, H., Li, Y., Shen, S., Pan, B., Wang, Q., Walker, H.C., Steffens, P., Boehm, M., et al. (2016). Evidence for a spinon Fermi surface in a triangular-lattice quantum-spin-liquid candidate. *Nature 540*, 559–562. 10.1038/nature20614.
28. Mo, Z., Jiang, J., Tian, L., Xie, H., Li, Y., Zheng, X., Zhang, L., Gao, X., Li, Z., and Liu, G. (2025). Ferromagnetic $Eu_2SiO_4$ compound with a record low-field magnetocaloric effect and excellent thermal conductivity near liquid helium temperature. *J. Am. Chem. Soc.* 147, 14684–14693. 10.1021/jacs.5c02997.
29. Xie, H., Tian, L., Zhang, L., Wang, J., Sun, H., Gao, X., Li, Z., Mo, Z., and Shen, J. (2023). Enhanced low-field magnetocaloric effect in Dy-doped hexagonal $GdBO_3$ compounds. *J. Rare Earths*. 41, 1728–1735. 10.1016/j.jre.2022.08.008.
30. Shu, M., Xu, X., Xi, N., He, M., Xiang, J., Qu, G., Khalyavin, D., Manuel, P., Nakamura, J.G., and Jiao, J. (2026). Giant magnetocaloric effect and spin supersolid in a metallic dipolar magnet. *Nature*

651, 61–67. 10.1038/s41586-026-10144-z.

31. Lorusso, G., Sharples, J.W., Palacios, E., Roubeau, O., Brechin, E.K., Sessoli, R., Rossin, A., Tuna, F., McInnes, E.J.L., Collison, D., and Evangelisti, M. (2013). A Dense Metal–Organic Framework for Enhanced Magnetic Refrigeration. *Adv. Mater.* 25, 4653–4656. 10.1002/adma.201301997.
32. Daudin, B., Lagnier, R., and Salce, B. (1982). Thermodynamic properties of the gadolinium gallium garnet, $Gd_3Ga_5O_{12}$, between 0.05 and 25 K. *J. Magn. Magn. Mater.* 27, 315–322. 10.1016/0304-8853(82)90092-0.
33. Paddison, J.A.M., Jacobsen, H., Petrenko, O.A., Fernández-Díaz, M.T., Deen, P.P., and Goodwin, A.L. (2015). Hidden order in spin-liquid $Gd_3Ga_5O_{12}$. *Science* 350, 179–181. 10.1126/science.aaa5326.
34. Wang, J., Fang, C., Qiu, Z., Zhao, Y., Xiao, Q., Sun, X., Li, Z., Li, L., Zhou, Y., Pan, C., and Guo, S. (2026). Tunable Multistage Refrigeration via Geometrically Frustrated Triangular Lattice Antiferromagnet for Space Cooling. *Device* 4, 101080. 10.1016/j.device.2026.101080.
35. Lin, W., Zhao, N., Li, Z., Zhao, Y., Liao, Y., An, W., Guo, R., Wang, J., Pan, C., Wen, B., et al. (2026). Quantum criticality enhanced millikelvin magnetic refrigeration in a large-spin-7/2 triangular lattice antiferromagnet. *Innovation* 7, 101254. 10.1016/j.xinn.2026.101254.
36. Tarasenko, M.S., Kiryakov, A.S., Ryadun, A.A., Kuratieva, N.V., Malyutina-Bronskaya, V.V., Fedorov, V.E., Wang, H.-C., and Naumov, N.G. (2022). Facile synthesis, structure, and properties of $Gd_2O_2Se$. *J. Solid State Chem.* 312. 123324. 10.1016/j.jssc.2022.123224.
37. Malyutina-Bronskaya, V.V., Soroka, A.S., Senkevich, D.V., Tarasenko, M.S., Ryadun, A.A., and Naumov, N.G. (2022). Optical Properties of Optically Pure Rare-Earth-Element Oxyselenides. *J. Appl. Spectrosc.* 89, 839–843. 10.1007/s10812-022-01434-1.
38. Quezel, G., Rossat-Mignod, J., and Lang, H.Y. (1972). Proprietes magnetiques des oxyseleniures de Gd, Tb, Dy, Ho, Er, Tm et structure magnetique de $Ho_2O_2Se$ et de $Yb_2O_2Se$. *Solid State Commun.* 10, 735–738. 10.1016/0038-1098(72)90182-2.
39. Abbas, Y., Rossat-Mignod, J., and Quezel, G. (1973). Magnetic structures and magnetic susceptibilities of terbium oxysulfide and oxyselenide. *Solid State Commun.* 12, 985–991. 10.1016/0038-1098(73)90021-5.
40. Stock, C., and McCabe, E.E. (2016). The magnetic and electronic properties of oxyselenides-influence of transition metal ions and lanthanides. *J Phys. Condens. Matter.* 28, 453001. 10.1088/0953-8984/28/45/453001.
41. Rutherford, A., Xing, C., Zhou, H., Huang, Q., Choi, E.S., and Calder, S. (2024). Magnetic properties of $R_2O_2CO_3$(R = Pr, Nd, Gd, Tb, Dy, Ho, Er, Yb) with a rare earth bilayer of triangular lattice. *Phys. Rev. Materials.* 8, 114413. 10.1103/PhysRevMaterials.8.114413.
42. Abbas, Y., Rossat-Mignod, J., Quezel, G., and Vettier, C (1974). Magnetic Structures of Dysprosium Oxysulfide and Dysprosium Oxyselenide. *Solid State Commun.* 14, 1115–1118. 10.1016/0038-1098(74)90285-3.
43. Abbas, Y., and , J.R.-M., G. Quezel (1973). Magnetic Structures and Magnetic Susceptibilities of Terbium Oxysulfide and Oxyselenide. *Solid State Commun.* 12, 985–991. 10.1016/0038-1098(73)90021-5.
44. Seabra, L., and Shannon, N. (2011). Competition between supersolid phases and magnetization plateaus in the frustrated easy-axis antiferromagnet on a triangular lattice. *Phys. Rev. B.* 83. 134412. 10.1103/PhysRevB.83.134412.
45. Ono, T., Tanaka, H., Aruga Katori, H., Ishikawa, F., Mitamura, H., and Goto, T. (2003). Magnetization plateau in the frustrated quantum spin system $Cs_2CuBr_4$. *Phys. Rev. B.* 67, 104431.

10.1103/PhysRevB.67.104431.
46. Quezel, G., Ballestracci, R., and Rossat-Mignod, J. (1970). Proprietes magnetiques des oxysulfures de terres rares. *J. Phys. Chem. Solids*. 31, 669–684. 10.1016/0022-3697(70)90201-5.
47. Yang, Y., Han, J., Zhou, Z., Zou, M., Xu, Y., Zheng, Y., Nan, C.W., and Lin, Y.H. (2022). Seeking New Layered Oxyselenides with Promising Thermoelectric Performance. *Adv. Funct. Mater.* 32. 10.1002/adfm.202113164.
48. Sheldrick, G. (2015). Crystal structure refinement with SHELXL. *Acta Crystallogr. C*. 71, 3–8. 10.1107/S2053229614024218.
49. Dolomanov, O.V., Bourhis, L.J., Gildea, R.J., Howard, J.A.K., and Puschmann, H. (2009). OLEX2: a complete structure solution, refinement and analysis program. *J. Appl. Crystallogr.* 42, 339–341. 10.1107/S0021889808042726.
50. CrystalMaker Software Ltd (2025). SingleCrystal 5 (CrystalMaker Software Ltd). https://www.crystalmaker.com/singlecrystal.

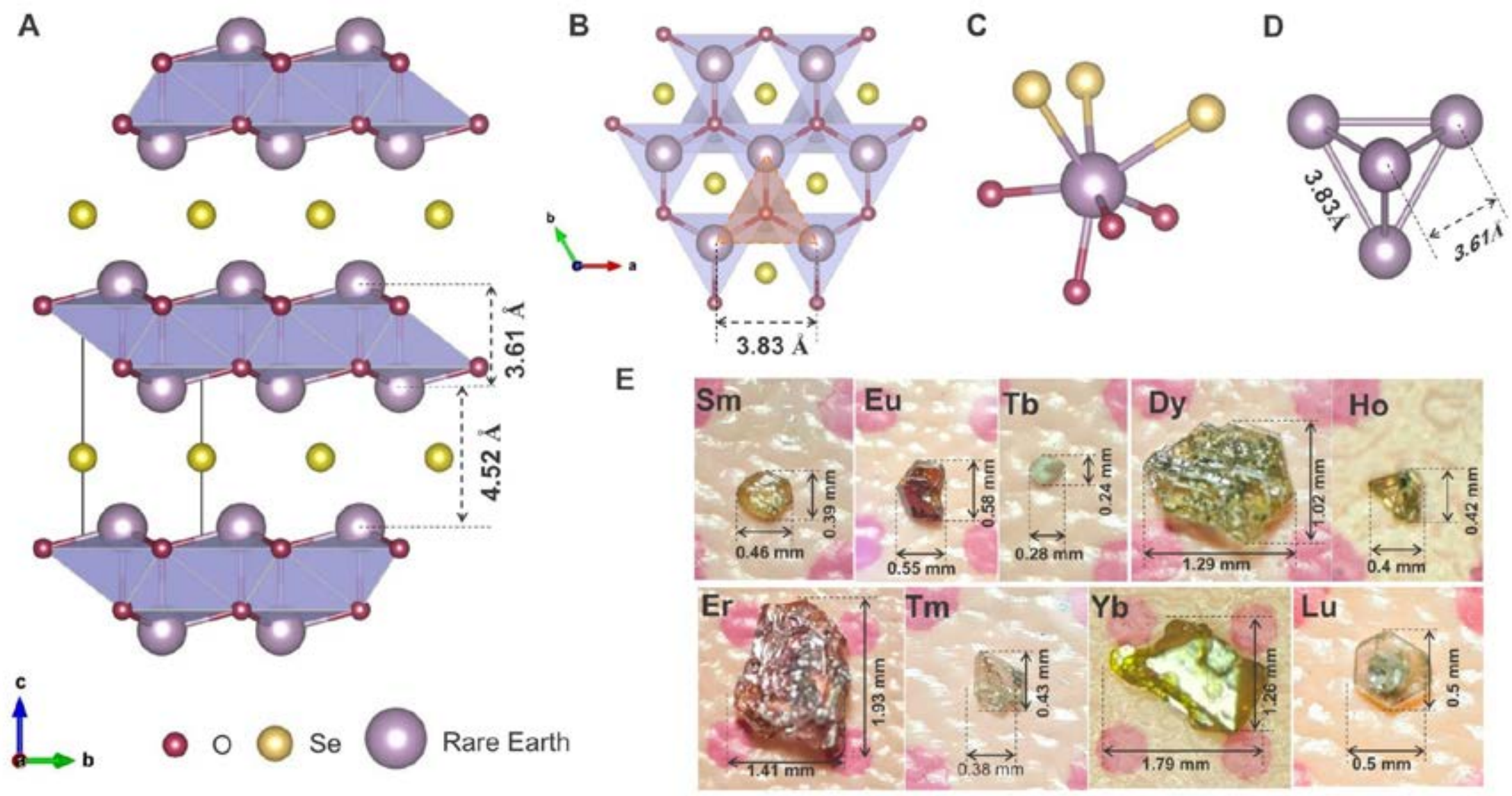

**Figure 1. Crystal morphology and structure**. Schematic illustration of the *R*OSe crystal structure along the (A) *a*-axis, (B) *c*-axis. (C) Coordination environment of $R^{3+}$ ions, (D) Schematic illustration of the tetrahedral unit formed by the double-layer $R^{3+}$, (E) *R*OSe crystals images. Bond distances shown correspond to the Dy-based compound.

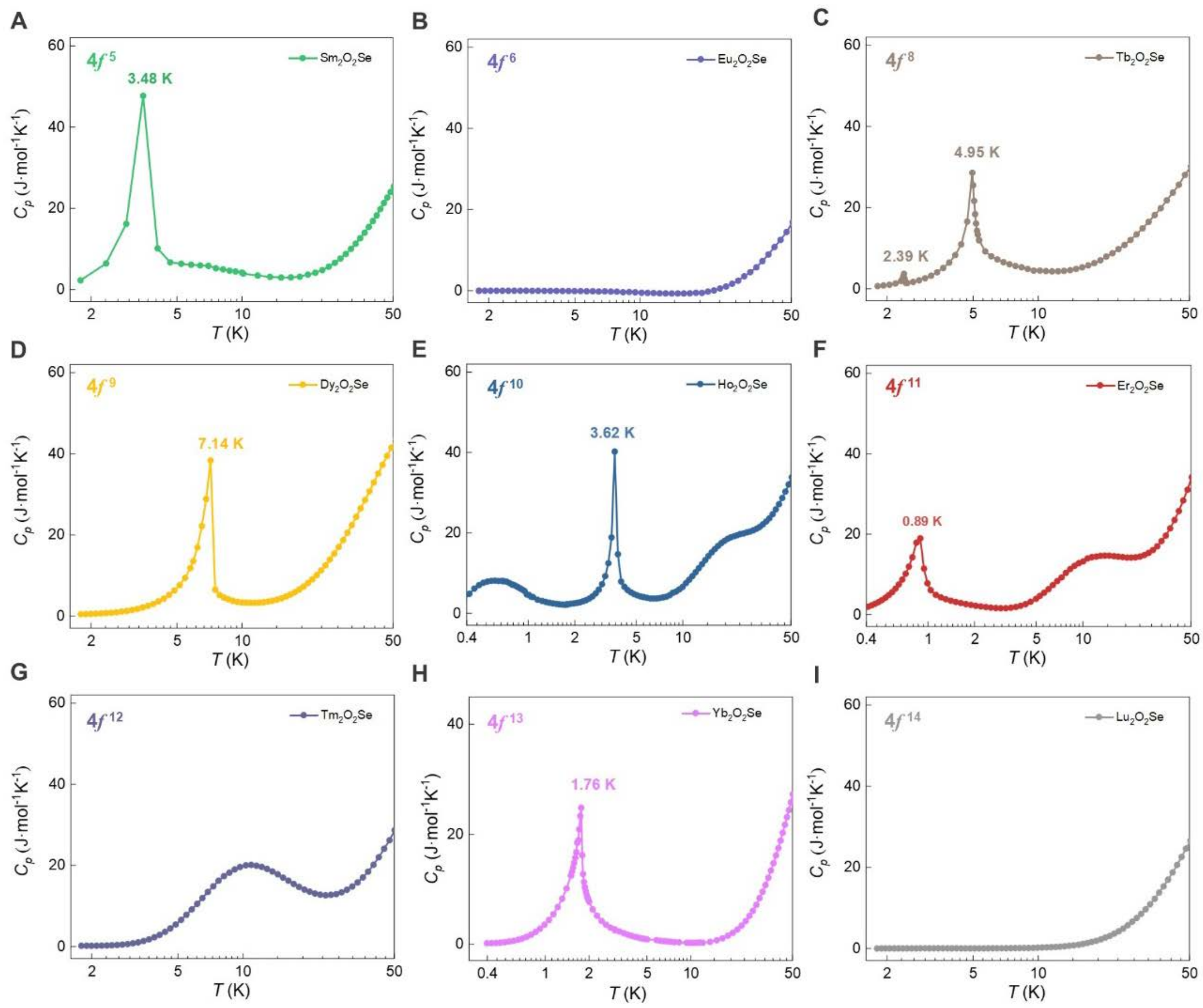

**Figure 2.** Zero-field heat capacity of *R*OSe below 50 K for *R* = (A) Sm, (B) Eu, (C) Tb, (D) Dy, (E) Ho, (F) Er, (G) Tm, (H) Yb, and (I) Lu.

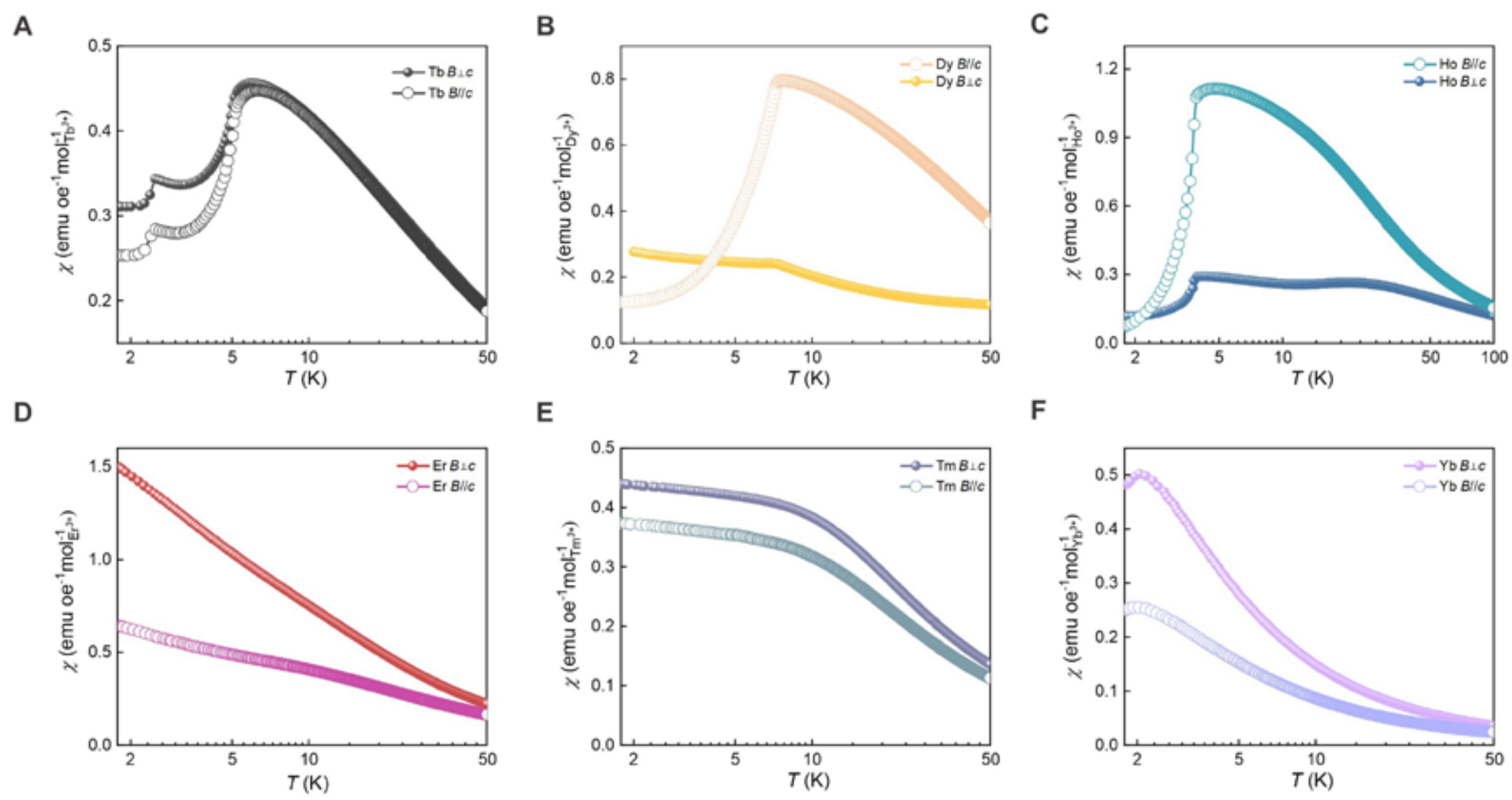

**Figure 3. Anisotropic temperature-dependent magnetic susceptibility of *R*OSe measured under an applied field of 0.05 T**. (A–F) *R* = Tb, Dy, Ho, Er, Tm, and Yb, respectively, with data collected under $B \perp c$ and $B // c$

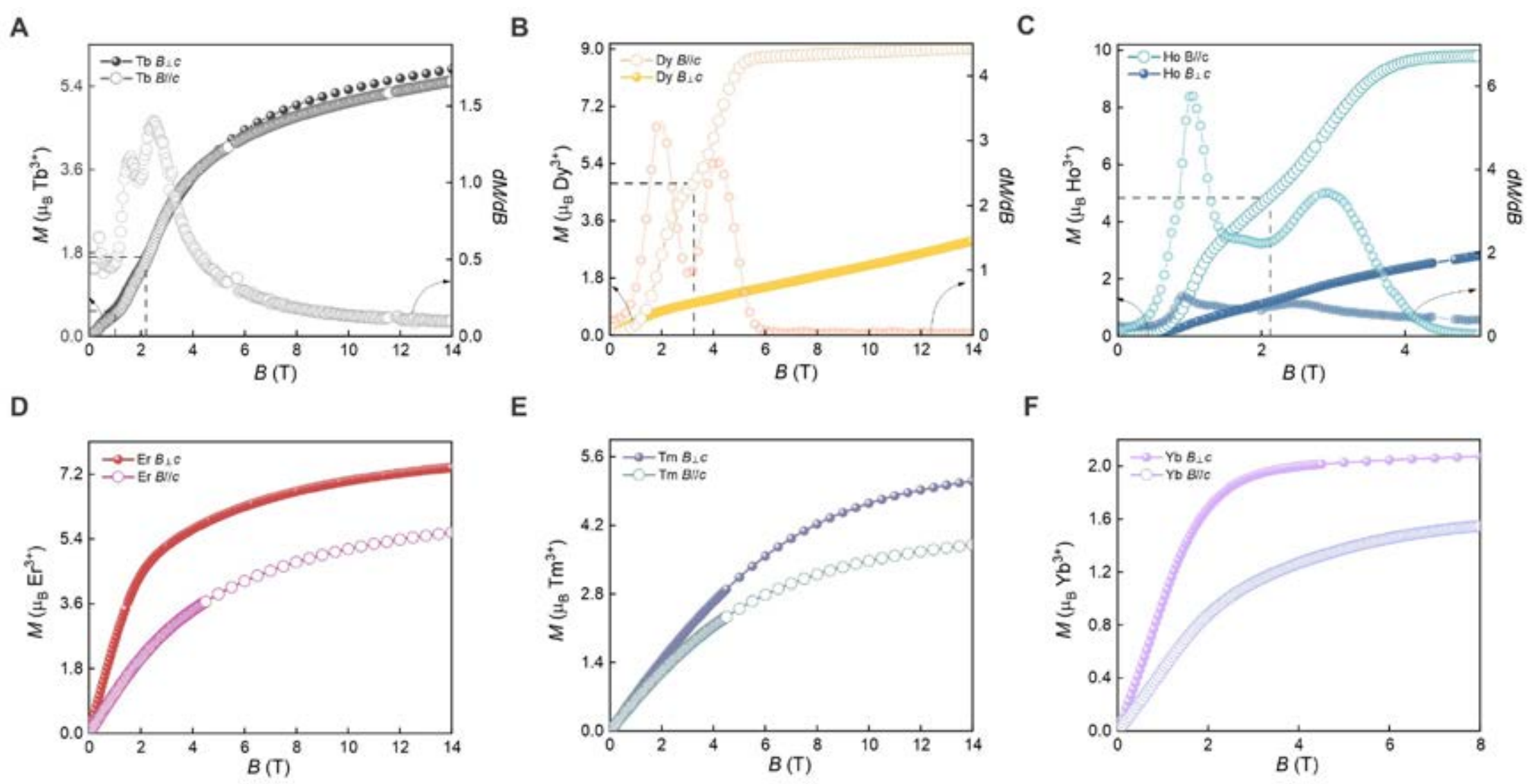

**Figure 4. Field-dependent magnetization of *R*OSe.** (A–F) $R$ = Tb, Dy, Ho, Er, Tm, and Yb, respectively, with data collected under $B \perp c$ and $B\ //\ c$.

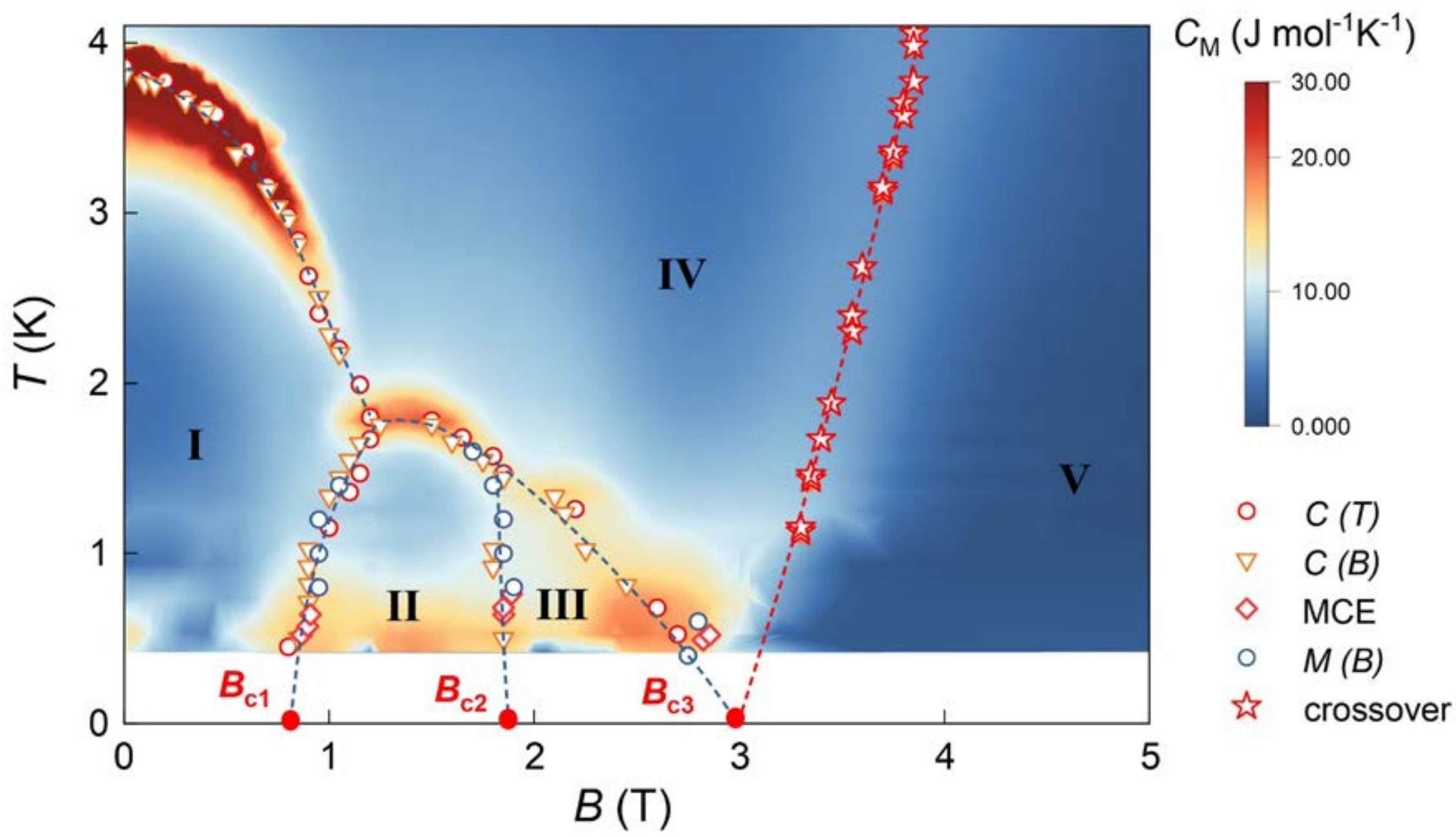

**Figure 5. Field-temperature magnetic phase diagram of HoOSe under *B* // *c*.** The magnetic phase boundaries are extracted from $M(B)$, $C_p(T)$, $C_p(B)$, and MCE measurements.

# Supporting information

**Table S1.** Crystallographic data of *R*OSe (*R* = Sm, Eu, Tb, Dy) series determined by single crystal X-ray diffraction at 100 K.

| Empirical formula | $Sm_2O_2Se$ | $Eu_2O_2Se$ | $Tb_2O_2Se$ | $Dy_2O_2Se$ |
|---|---|---|---|---|
| Space group | *P*-3*m*1 | *P*-3*m*1 | *P-3m1* | *P-3m1* |
| Formula weight | 411.66 | 414.88 | 428.8 | 435.6 |
| Temperature/K | 100 | 100 | 100 | 100 |
| *a*/Å | 3.9223(3) | 3.9017(2) | 3.8554(2) | 3.8321(2) |
| *b*/Å | 3.9223(3) | 3.9017(2) | 3.8554(2) | 3.8321(2) |
| *c*/Å | 6.9092(7) | 6.8761(7) | 6.8146(7) | 6.7864(5) |
| $\alpha$/° | 90 | 90 | 90 | 90 |
| $\beta$/° | 90 | 90 | 90 | 90 |
| $\gamma$/° | 120 | 120 | 120 | 120 |
| Cell volume /$Å^3$ | 92.053(17) | 90.653(13) | 87.722(13) | 86.307(11) |
| Z | 1 | 1 | 1 | 1 |
| Density (calculated) /$g·cm^{-3}$ | 7.426 | 7.600 | 8.117 | 8.388 |
| Absorption coefficient /$mm^{-1}$ | 41.253 | 44.097 | 50.130 | 53.269 |
| F (000) | 174 | 176 | 180 | 182 |
| Crystal size /$mm^3$ | 0.092×0.034×0.033 | 0.119×0.054×0.045 | 0.035×0.031×0.017 | 0.057×0.041×0.035 |
| Radiation/Å | Mo $K_\alpha$ (0.71073) | | | |
| Theta range for data collection/° | 5.896 to 60.482 | 5.924 to 60.832 | 5.978 to 60.562 | 6.004 to 70.368 |
| Reflections collected | 3577 | 3513 | 3328 | 3351 |
| Independent reflections | 134 | 134 | 130 | 178 |
| Independent reflections R indices | 0.0640 | 0.0617 | 0.0524 | 0.0347 |
| Goodness-of-fit on $F^2$ | 1.259 | 1.358 | 1.220 | 1.185 |
| Final R indices [$I \geq 2\sigma(I)$] | $R_1$=0.0128<br>$wR_2$=0.0298 | $R_1$=0.0129<br>$wR_2$=0.0253 | R1=0.0124<br>wR2=0.0263 | R1=0.0166<br>wR2=0.0401 |
| R indices [all data] | $R_1$=0.0129<br>$wR_2$=0.0298 | $R_1$=0.0145<br>$wR_2$=0.0256 | R1=0.0125<br>wR2=0.0263 | R1=0.0167<br>wR2=0.0402 |

**Table S2.** Crystallographic data of *R*OSe (*R* = Ho, Er, Tm, Yb, Lu) series determined by single crystal X-ray diffraction at 100 K.

| Empirical formula | $Ho_2O_2Se$ | $Er_2O_2Se$ | $Tm_2O_2Se$ | $Yb_2O_2Se$ | $Lu_2O_2Se$ |
|---|---|---|---|---|---|
| Space group | *P*-3*m*1 | *P*-3*m*1 | *P*-3*m*1 | *P*-3*m*1 | *P*-3*m*1 |
| Formula weight | 440.82 | 445.48 | 448.82 | 457.04 | 460.90 |
| Temperature/K | 100 | 100 | 100 | 100 | 100 |
| *a*/Å | 3.8140(2) | 3.7925(3) | 3.7744(2) | 3.7542(3) | 3.7409(2) |
| *b*/Å | 3.8140(2) | 3.7925(3) | 3.7744(2) | 3.7542(3) | 3.7409(2) |
| *c*/Å | 6.7622(5) | 6.7356(7) | 6.7091(7) | 6.6753(11) | 6.6738(6) |
| $\alpha$/° | 90 | 90 | 90 | 90 | 90 |
| $\beta$/° | 90 | 90 | 90 | 90 | 90 |
| $\gamma$/° | 120 | 120 | 120 | 120 | 120 |
| Cell volume /$Å^3$ | 85.188(12) | 83.899(16) | 82.773(12) | 81.477(19) | 80.883(11) |
| Z | 1 | 1 | 1 | 1 | 1 |
| Density (calculated) /$g{\cdot}cm^{-3}$ | 8.593 | 8.817 | 9.004 | 9.315 | 9.462 |
| Absorption coefficient /$mm^{-1}$ | 56.551 | 60.281 | 64.000 | 67.964 | 71.678 |
| F (000) | 184 | 186 | 188 | 190 | 192 |
| Crystal size /$mm^3$ | 0.114×0.061×0.059 | 0.085×0.052×0.047 | 0.119×0.106×0.05 | 0.088×0.055×0.02 | 0.086×0.0041×0.035 |
| Radiation/Å | | | Mo K$\alpha$ (0.71073) | | |
| Theta range for data collection/° | 6.024 to 54.756 | 6.048 to 60.918 | 6.072 to 60.614 | 6.104 to 60.966 | 6.104 to 70.298 |
| Reflections collected | 2607 | 2027 | 2643 | 1541 | 3721 |
| Independent reflections | 93 | 127 | 122 | 123 | 164 |
| Independent reflections R indices | 0.0500 | 0.0916 | 0.0568 | 0.0661 | 0.0469 |
| Goodness-of-fit on $F^2$ | 1.203 | 1.194 | 1.212 | 1.189 | 1.162 |
| Final R indices [$I \geq 2\sigma(I)$] | $R_1$=0.0181<br>$wR_2$=0.0434 | $R_1$=0.0187<br>$wR_2$=0.0398 | $R_1$=0.0170<br>$wR_2$=0.0422 | $R_1$=0.0168<br>$wR_2$=0.0325 | $R_1$=0.0167<br>$wR_2$=0.0394 |
| R indices [all data] | $R_1$=0.0181<br>$wR_2$=0.0434 | $R_1$=0.0209<br>$wR_2$=0.0400 | $R_1$=0.0173<br>$wR_2$=0.0422 | $R_1$=0.0192<br>$wR_2$=0.0328 | $R_1$=0.0170<br>$wR_2$=0.0397 |

**Table S3.** Magnetic ordering temperatures and characteristic magnetic parameters.

| **Materials** | **$T_N$ (K)** | **$M_{max}$** ($\mu_B/R^{3+}$) | **$\mu_{eff}$** ($\mu_B/R^{3+}$) | $\Theta_{cw}$ (K) |
|---|---|---|---|---|
| | | | (Fitting range: 100-300 K) | |
| $Sm_2O_2Se$ | 3.48 | (0.71) | -- | -- |
| $Tb_2O_2Se$ | 4.95<br>2.39 | 5.77 $B$⊥c (14 T)<br>4.54 $B$//c (7 T)<br>(9) | 10.08 $B$⊥c<br>9.89 $B$//c<br>(9.72) | -16.13 $B$⊥c<br>-14.10 $B$//c |
| $Dy_2O_2Se$ | 7.14 | 2.97 $B$⊥c (14 T)<br>8.80 $B$//c (7 T)<br>(10) | 10.89 $B$⊥c<br>10.58 $B$//c<br>(10.64) | -63.00 $B$⊥c<br>19.26 $B$//c |
| $Ho_2O_2Se$ | 3.62 | 7.00 $B$⊥c (14 T)<br>9.80 $B$//c (5 T)<br>(10) | 10.07 $B$⊥c<br>11.02 $B$//c<br>(10.61) | -11.64 $B$⊥c<br>10.89 $B$//c |
| $Er_2O_2Se$ | / | 7.38 $B$⊥c (14 T)<br>2.79 $B$//c (7 T)<br>(9) | 9.45 $B$⊥c<br>8.83 $B$//c<br>(9.58) | -1.02 $B$⊥c<br>-10.78 $B$//c |
| $Tm_2O_2Se$ | / | 10.21 $B$⊥c (14 T)<br>3.8 $B$//c (7 T)<br>(7) | 7.65 $B$⊥c<br>7.07 $B$//c<br>(7.56) | -2.49 $B$⊥c<br>-8.09 $B$//c |
| $Yb_2O_2Se$ | 1.76 | 4.24 $B$⊥c (14 T)<br>1.51 $B$//c (7T)<br>(4) | 4.92 $B$⊥c<br>4.42 $B$//c<br>(4.54) | -43.87 $B$⊥c<br>-79.70 $B$//c |

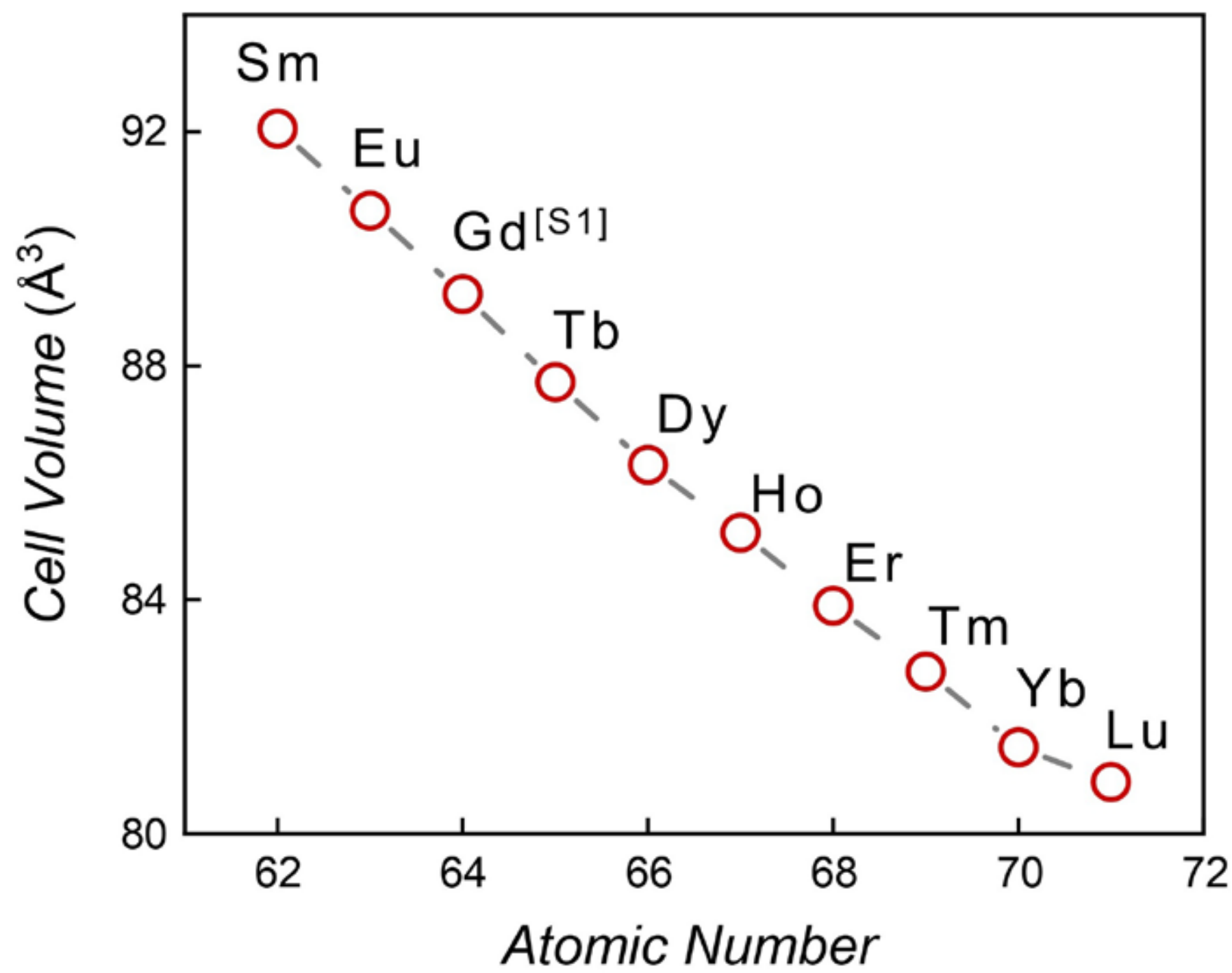

**Figure S1.** The lanthanide contraction effect on *R*OSe.

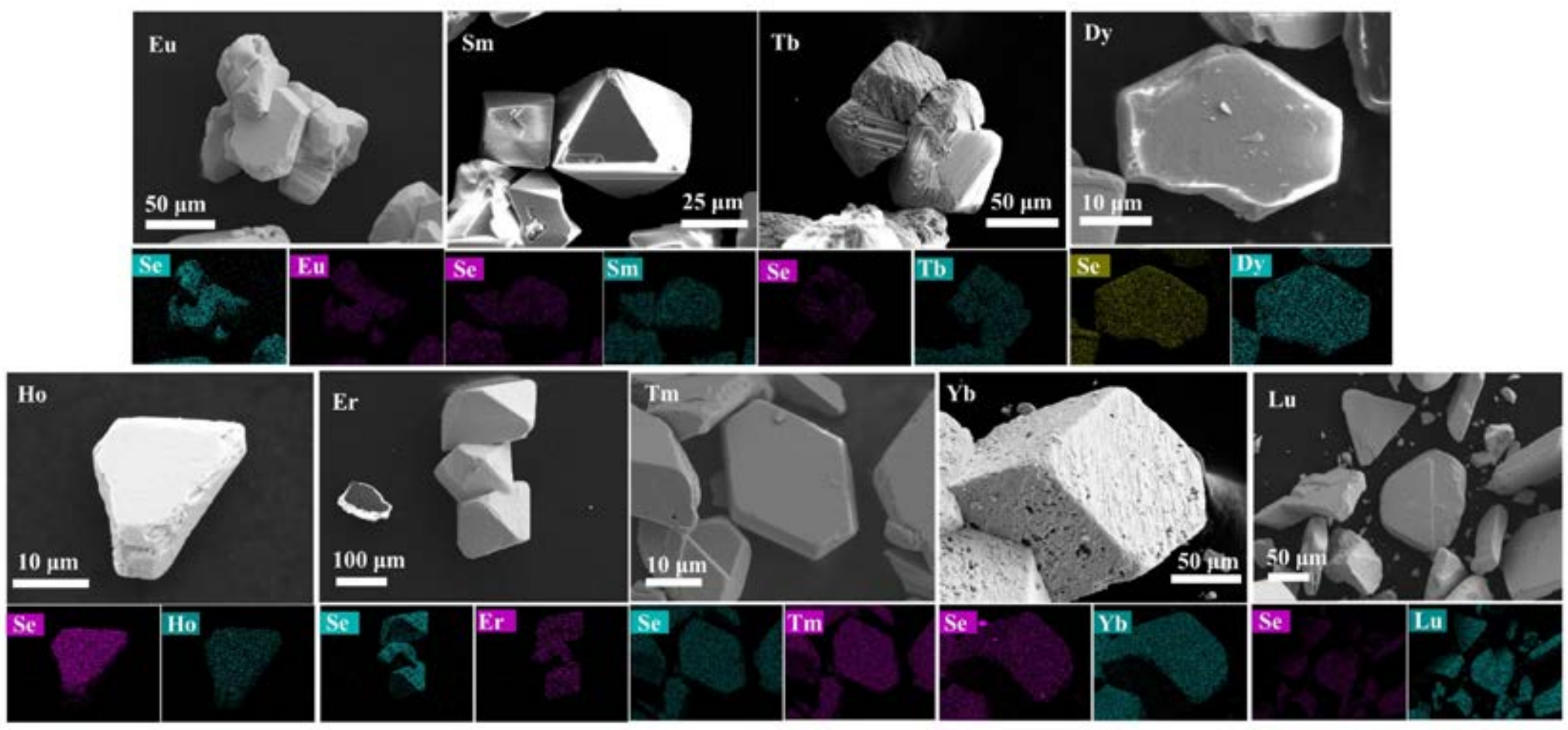

**Figure S2.** The SEM image of *R*OSe (*R* = Sm-Eu and Tb-Lu) single crystal, the distribution of rare-earth and Se elements on the surface of *R*OSe was determined by EDS.

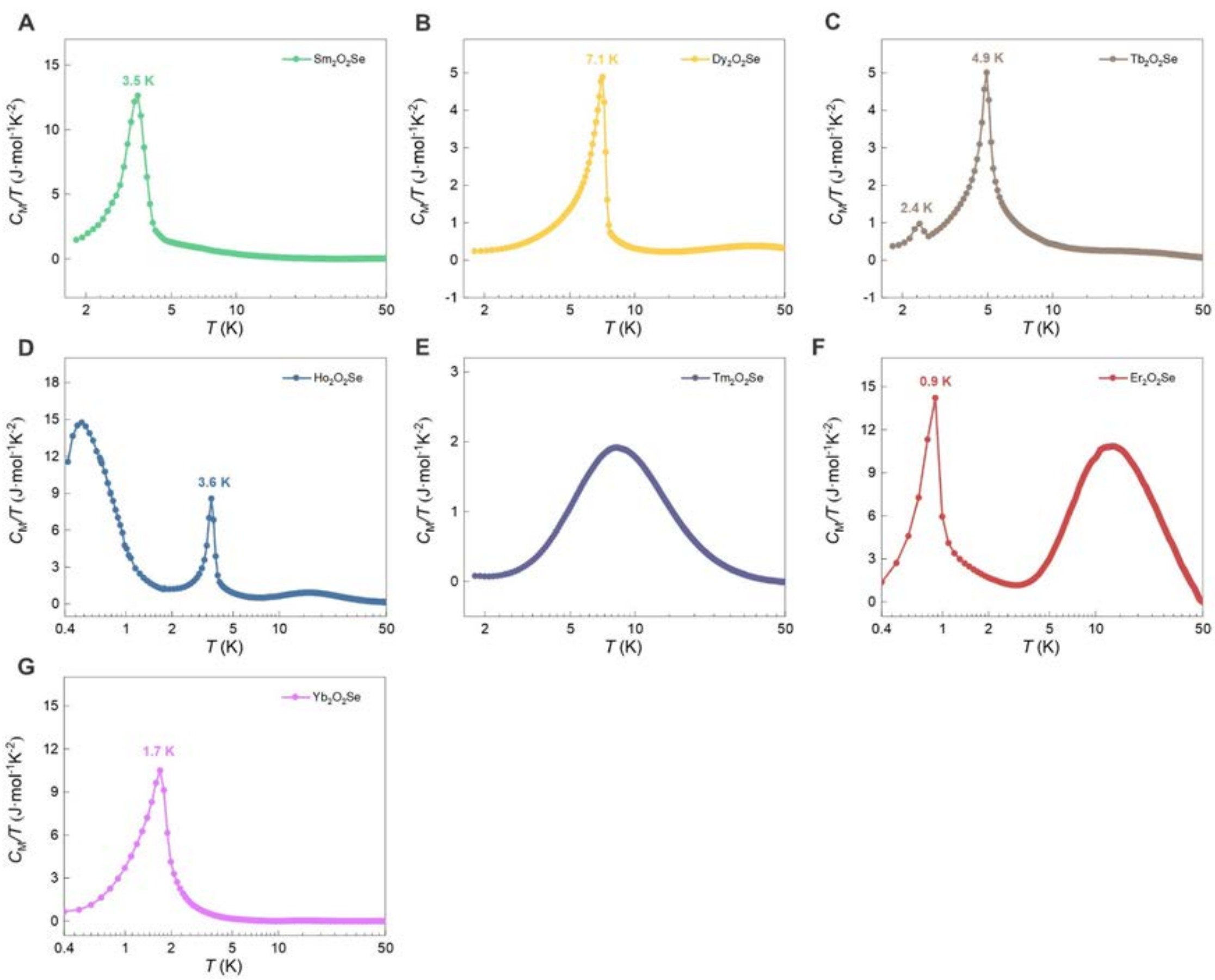

**Figure S3.** Magnetic specific heat $C_M/T$ of *R*OSe below 50 K for *R* = (A) Sm, (B) Dy, (C) Tb, (D) Ho, (E) Tm, (F) Er, (G) Yb.

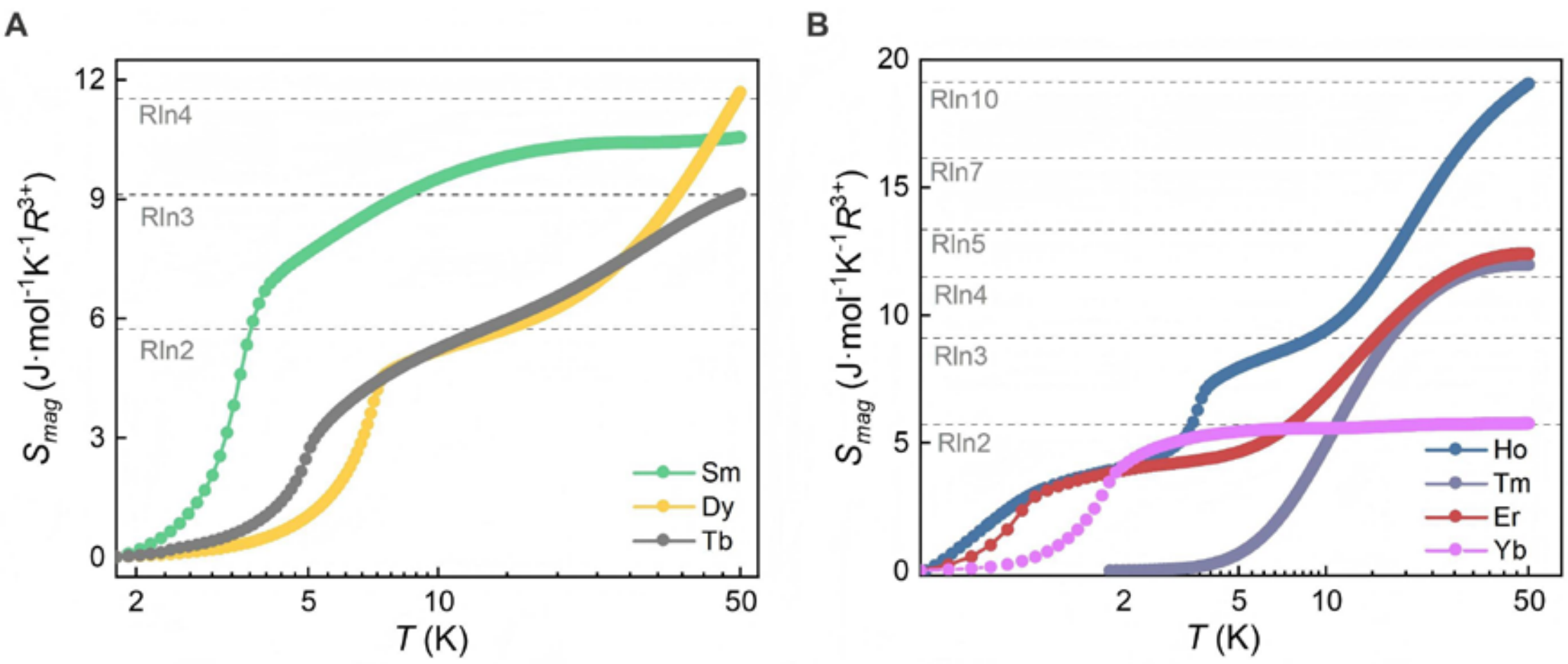

**Figure S4.** Magnetic entropy for (A) $R$ = Sm–Dy (excluding Gd) and (B) $R$ = Ho–Lu.

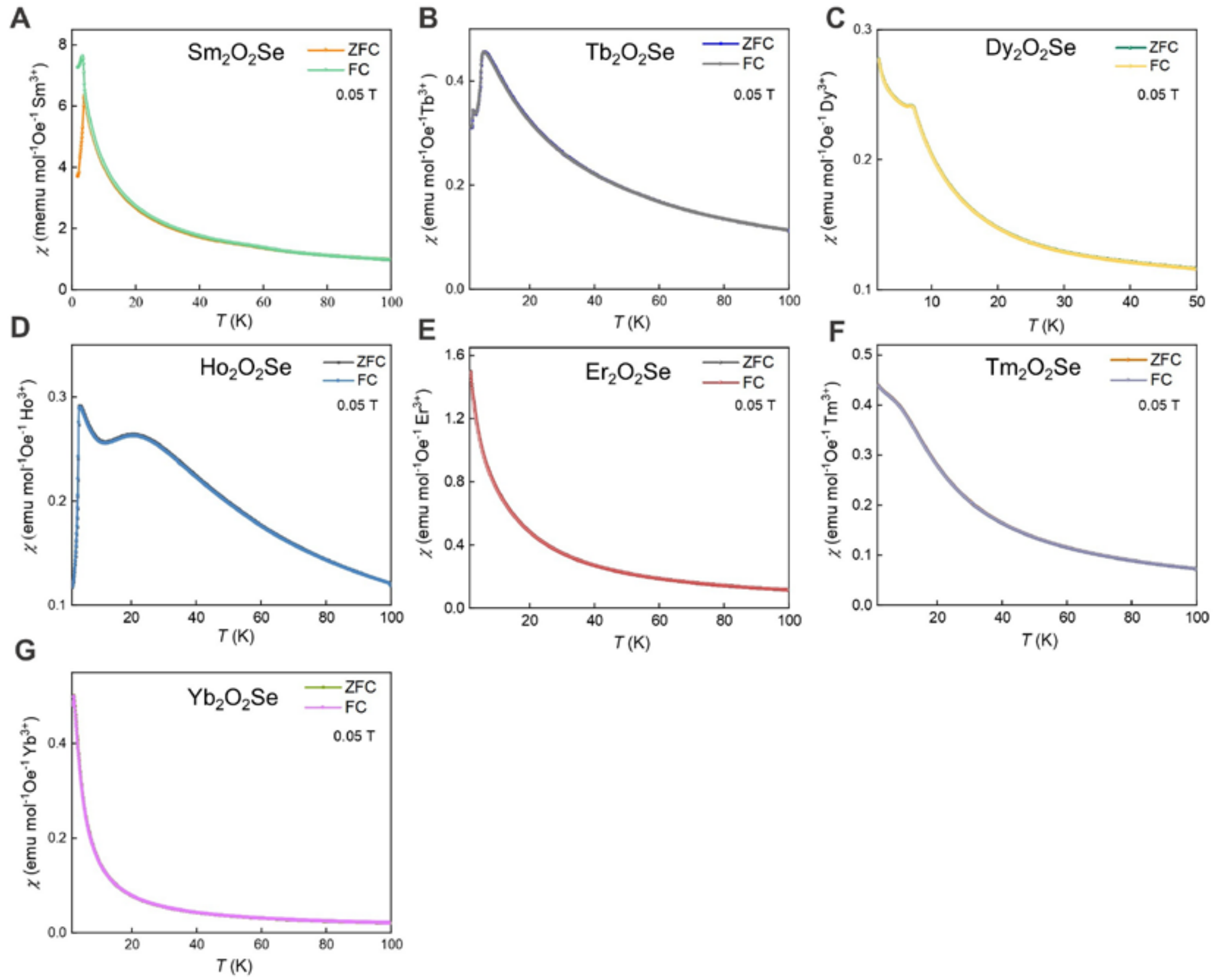

**Figure S5.** The ZFC and FC curves of *R*OSe for $B \perp c$.

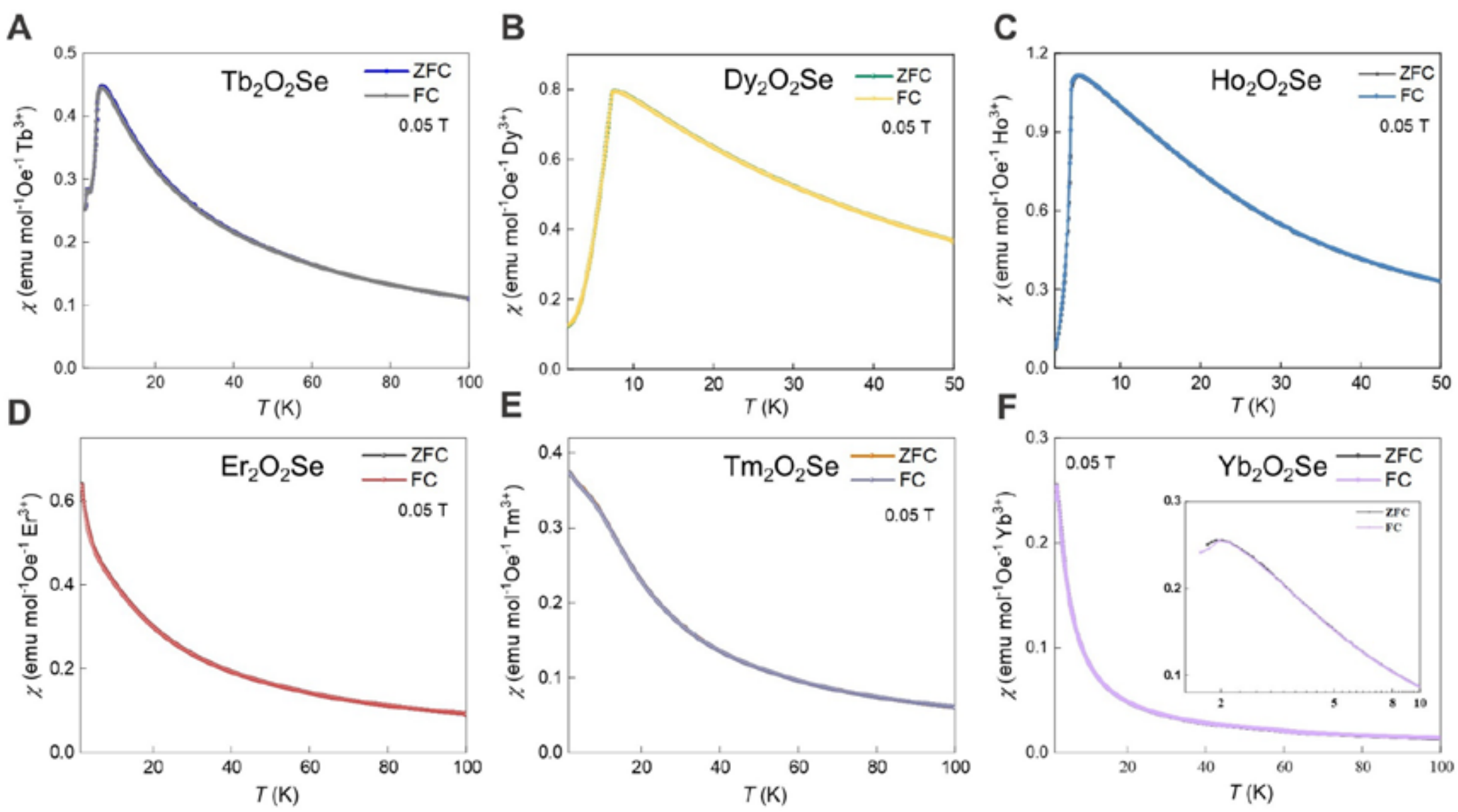

**Figure S6.** The ZFC and FC curve of *R*OSe for *B* // *c*.

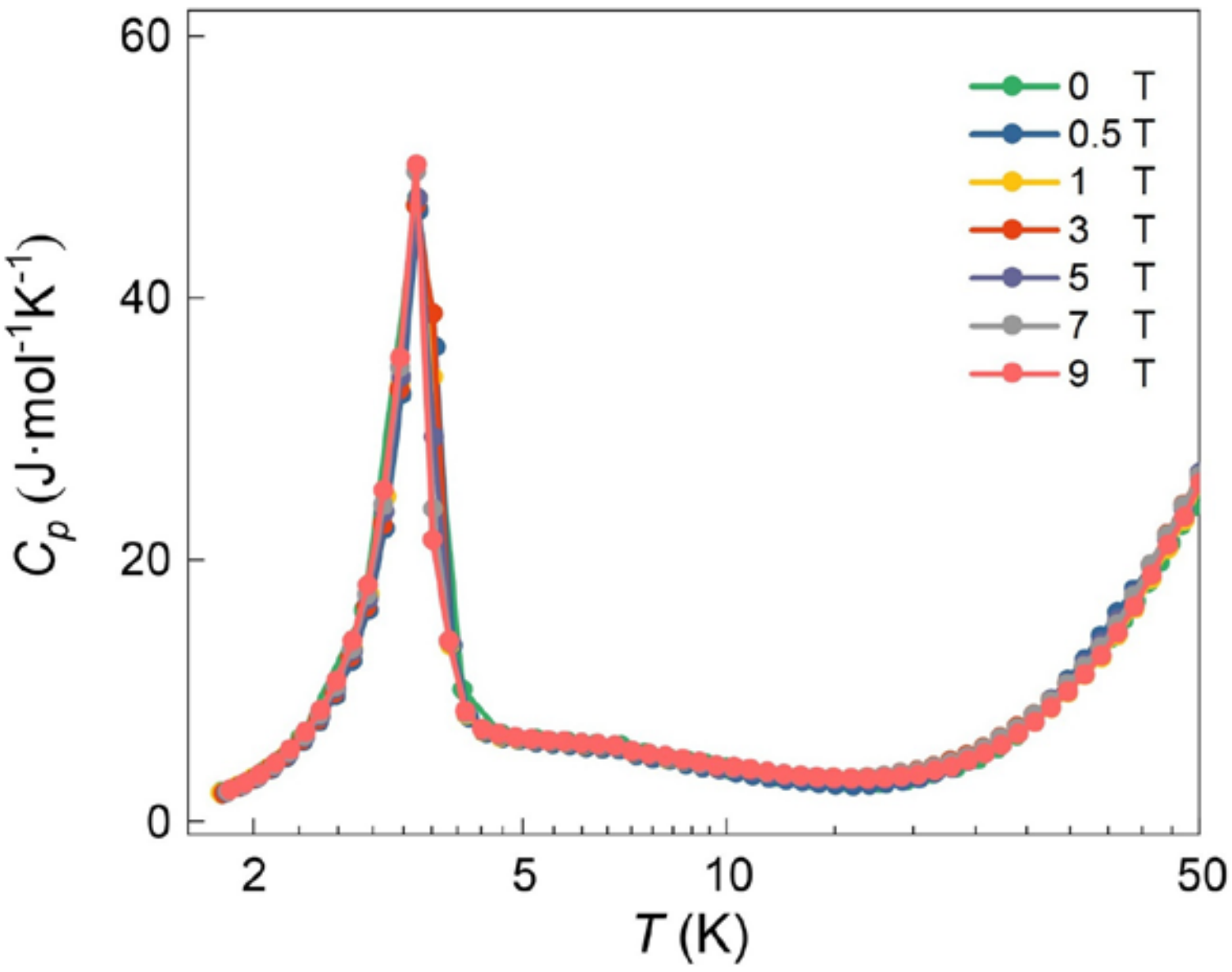

**Figure S7.** Field-dependent specific heat for $Sm_2O_2Se$ with applied fields $B$ // $c$.

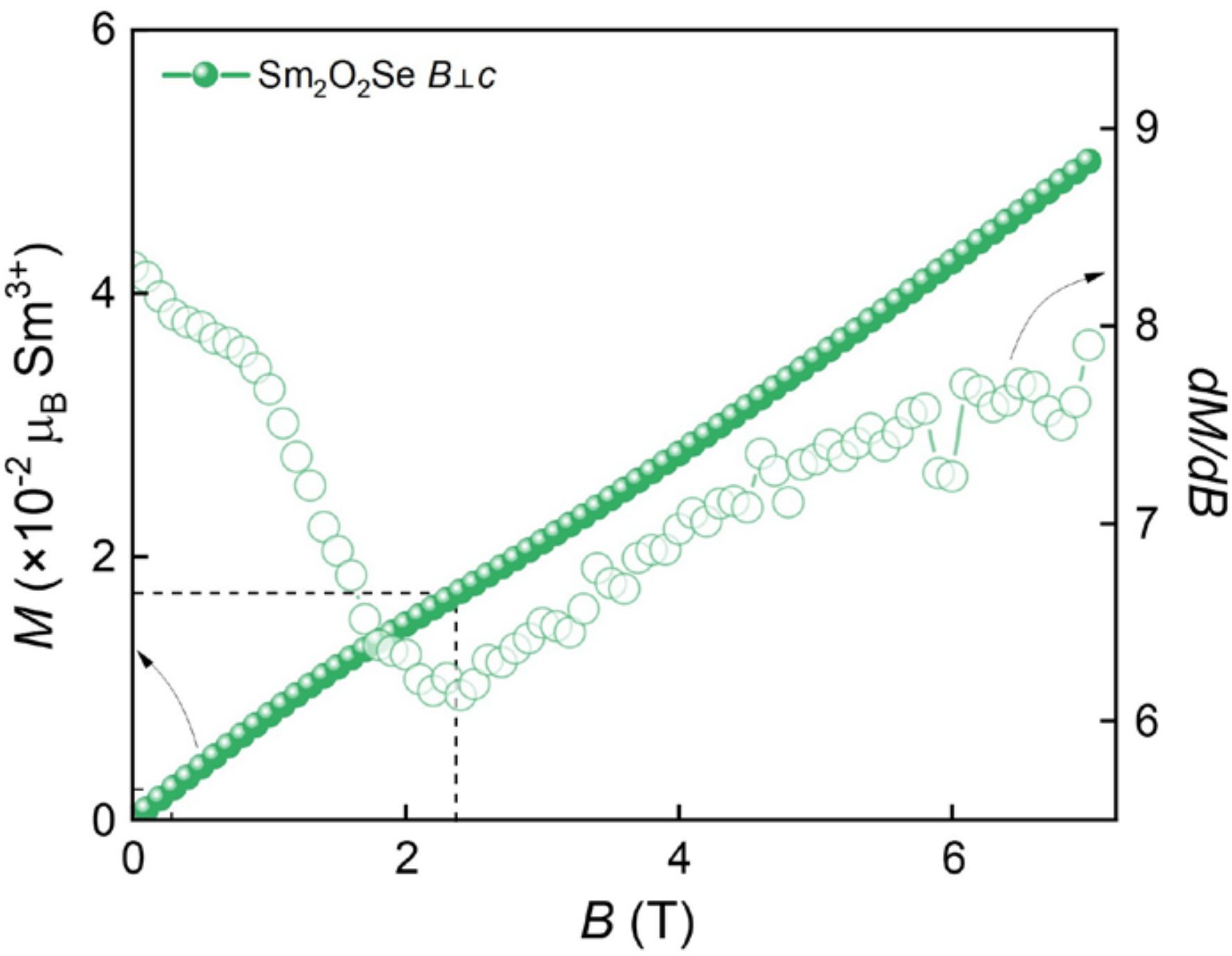

**Figure S8.** Field-dependent magnetization of $Sm_2O_2Se$.

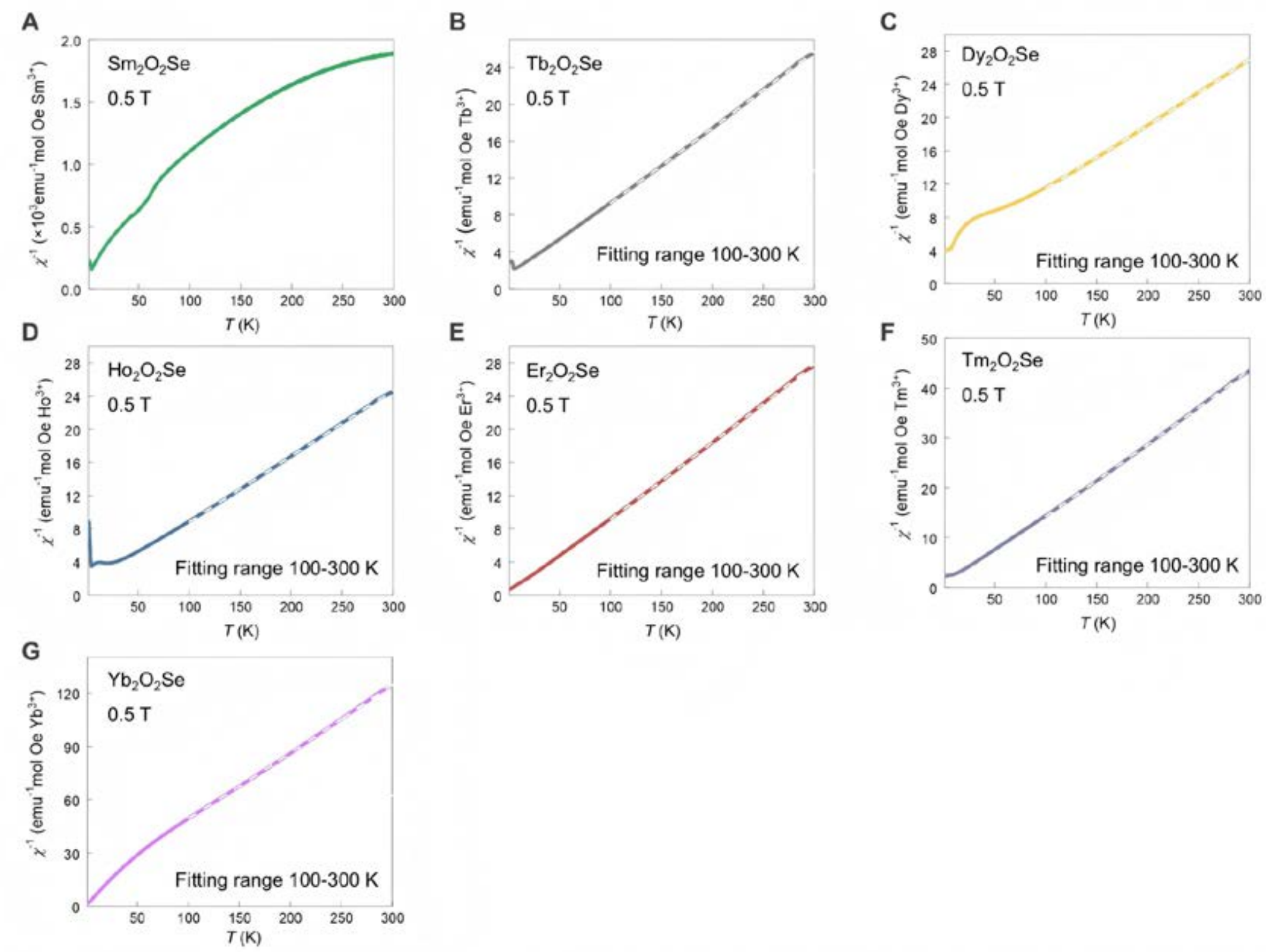

**Figure S9.** Inverse magnetic susceptibility as a function of temperature under a magnetic field of 0.5 T for $B \perp c$.

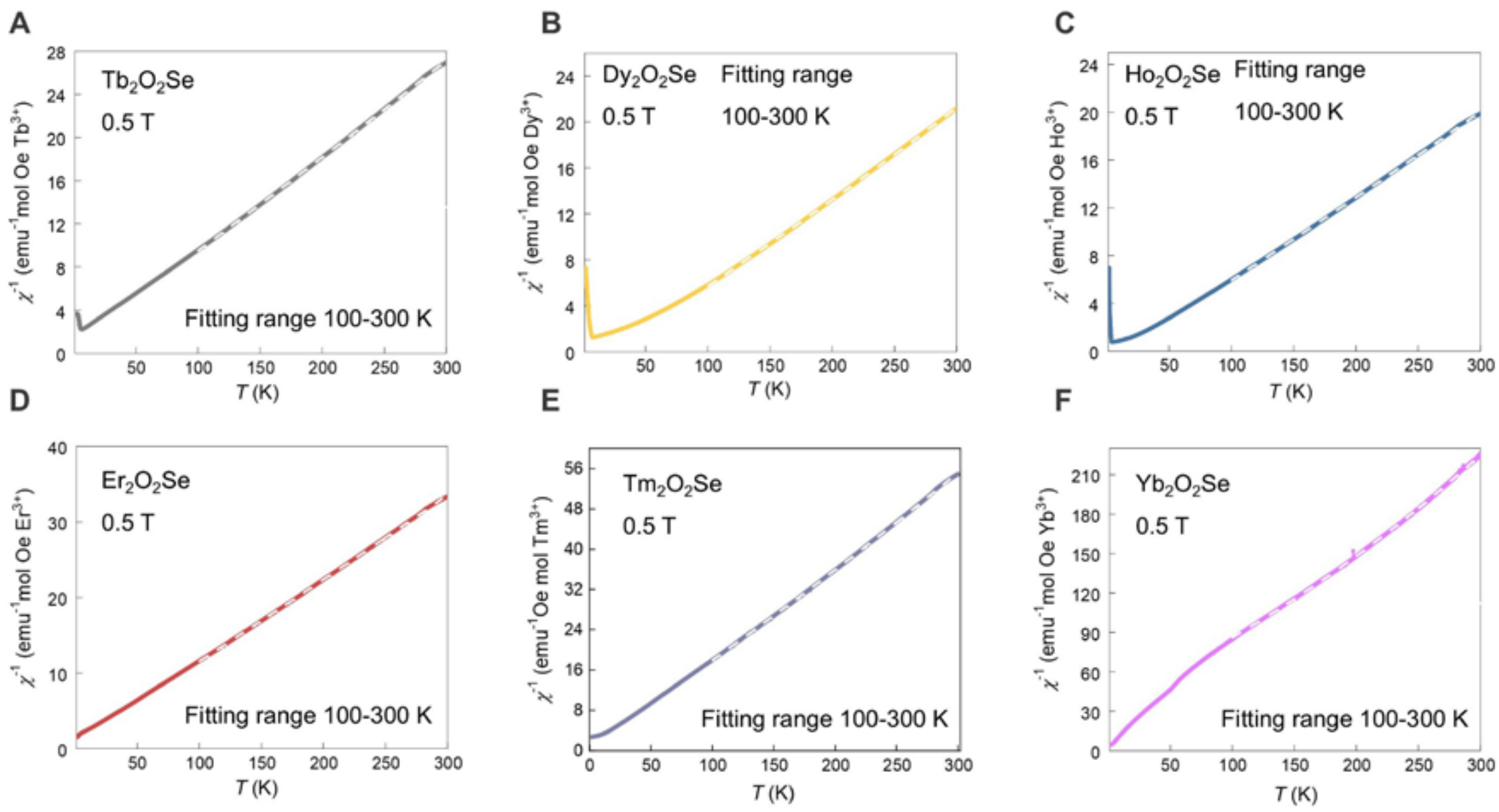

**Figure S10.** Inverse magnetic susceptibility as a function of temperature under a magnetic field of 0.5 T for *B* // c.

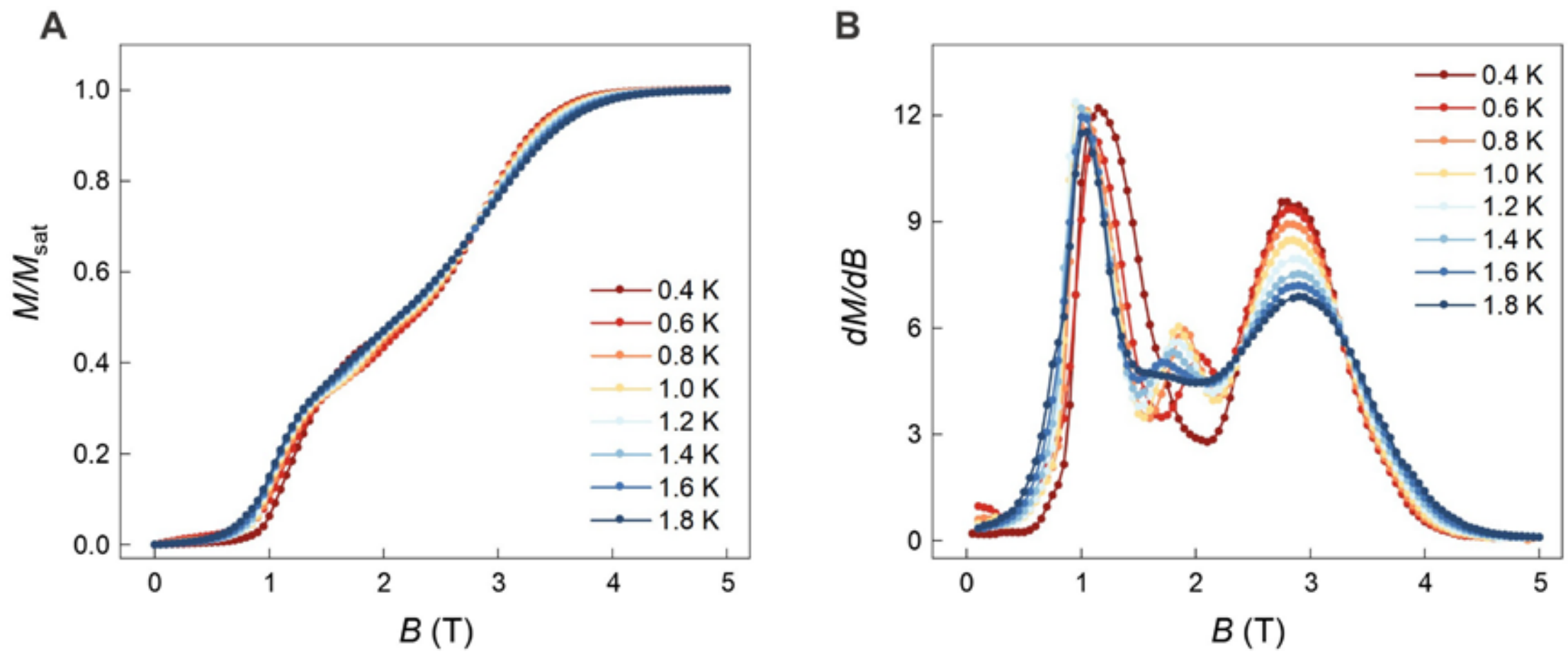

**Figure S11. Low-temperature field-dependent magnetization of $Ho_2O_2Se$**. (A) Normalized magnetization $M/M_{sat}$ measured between 0.4 and 1.8 K. (B) Corresponding $dM/dB$ curves, revealing multiple field-induced magnetic anomalies.

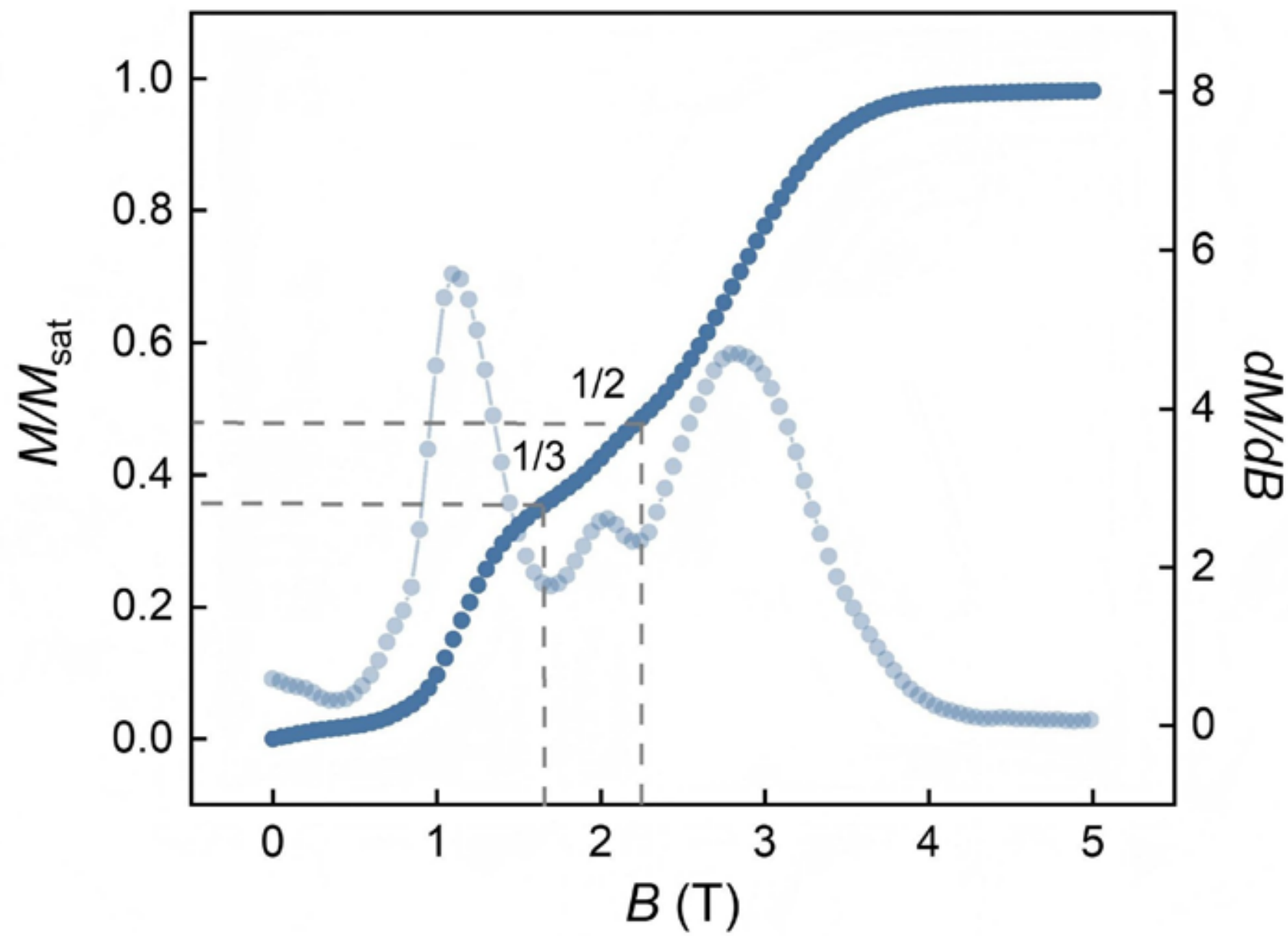

**Figure S12. Low-temperature field-dependent magnetization of $Ho_2O_2Se$ at 0.6 K.**

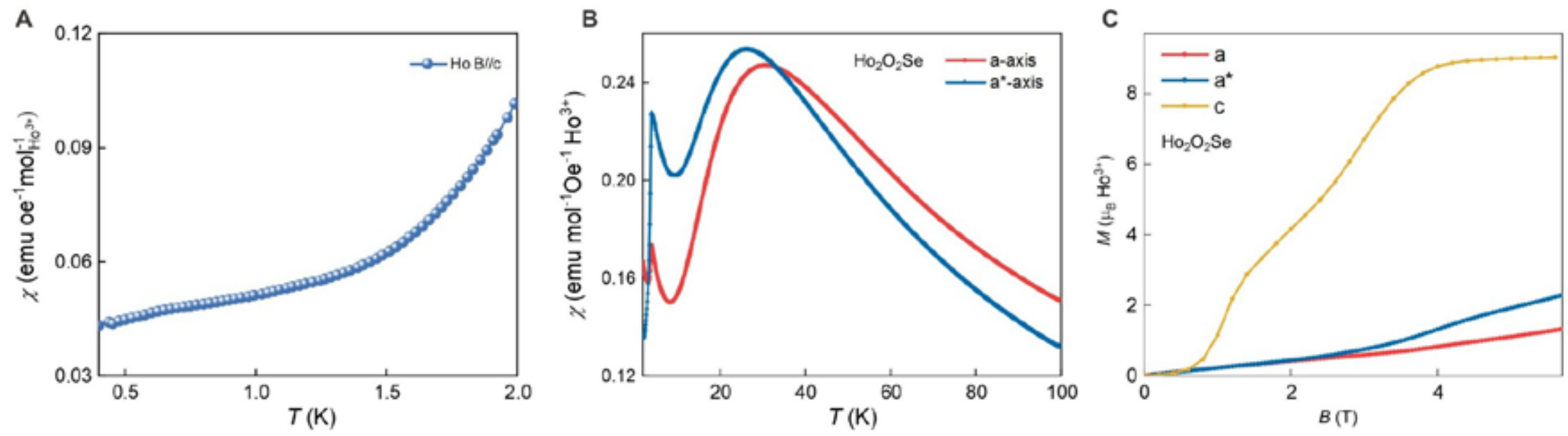

**Figure S13. Anisotropic magnetic properties of $Ho_2O_2Se$.** (A) Low-temperature magnetic susceptibility for *B* // *c* measured from 0.4 to 2.0 K under the field of 0.05 T. (B) Temperature-dependent susceptibility with field applied along the *a*-axis and $a^*$-axis. (C) Isothermal magnetization versus magnetic field *B* along the *a*-, *a**-, and *c*-axes.

**Reference**

S1. Wang, J., Fang, C., Qiu, Z., Zhao, Y., Xiao, Q., Sun, X., Li, Z., Li, L., Zhou, Y., Pan, C., and Guo, S. (2026). Tunable Multistage Refrigeration via Geometrically Frustrated Triangular Lattice Antiferromagnet for Space Cooling. Device 4, 101080. 10.1016/j.device.2026.101080.